# HYDROPHOBIC POLYELECTROLYTES IN BETTER POLAR SOLVENT. STRUCTURE AND CHAIN CONFORMATION AS SEEN BY SAXS AND SANS ‡

Wafa ESSAFI[1,*], Marie-Noelle SPITERI[2], Claudine WILLIAMS[3] and François BOUE[2]

**1** Institut National de Recherche et d'Analyse Physico-Chimique, Pôle Technologique de Sidi Thabet, 2020 Sidi Thabet – Tunisie

**2** Laboratoire Léon Brillouin, UMR 12 CNRS- IRAMIS CEA Saclay, 91191 Gif sur Yvette Cedex, France

**3** Physique de la Matière Condensée, UMR 7125 CNRS - Collège de France, 11 Place Marcelin Berthelot, F 75321 Paris, France

**Abstract**

We demonstrate in this paper the influence of solvent quality on the structure of the semi-dilute solution of a hydrophobic polyelectrolyte, partially sulfonated Poly-Styrene Sulfonate. Two solvents are used: (i) one mixture of water and an organic solvent: THF, which is also slightly polar; (ii) DMSO, a polar organic solvent. In case (i), it is shown by SAXS study that the structure - namely the scattering from all chains, characterised by a maximum ("polyelectrolyte peak"), of the aqueous hydrophobic polyelectrolyte solutions (PSS) depends on the solvent quality through the added amount of organic solvent THF. This dependence is more pronounced when the sulfonation rate is low (more hydrophobic polyelectrolyte).  It is proposed that when THF is added, the chain conformation evolves from the pearl necklace shape already reported in pure water, towards the conformation in pure water for fully sulfonated PSS, which is string-like as also reported previously. On the contrary, for a hydrophilic polyelectrolyte, AMAMPS, no evolution occurs with added THF in the aqueous solution. In case (ii), it is shown directly by SANS study that PSS can behave as a classical solvophilic polyelectrolyte when dissolved in an organic polar solvent such as DMSO: the structure (total scattering) as well as the form factor (single chain scattering measured by SANS using the Zero Average Contrast method) of the PSS chains is independent of the charge content in agreement with Manning condensation, and identical to the one of a fully charged PSS chain in pure water, which has a classical polyelectrolyte behaviour in the semi-dilute regime.

Poly(styrene)-co-(styrene sulfonate, sodium salt)
Poly(acrylamido – methyl propane sulfonate)

**I-INTRODUCTION**

Polyelectrolytes are polymers bearing ionisable groups, i. e. in presence of suitable polar solvent, these groups dissociate into electrostatically charged groups linked to the polymer backbone, and mobile counter-ions. Polyelectrolytes are called "hydrophobic" when ionisation coexists with attractive forces in water (we consider here attraction due to hydrophobicity only). In other words, the process of their solvation is then a combination of electrostatic and hydrophobic interactions. This is due to some parts of the chains for which water is a bad solvent (backbone or part of the backbone, short grafted entities, copolymers with one hydrophobic monomer). Conversely, all parts (ionisable or not) of "hydrophilic polyelectrolytes" are under good solvent conditions in water; the attractive contribution is negligible and interactions have mainly a pure electrostatic nature (essentially repulsive between two parts of chain).

The first category of polymers i.e. hydrophobic polyelectrolytes can however be ionised in a different polar solvent where at the same time the attractive contributions are lower or negligible. In this paper, we will examine the solution structure and the chain conformation under such conditions, in comparison with the case of pure water.

Let us recall the general background about polyelectrolytes chain conformation and interactions in a solvent. Let us start by the **dilute regime**. When water is a good solvent (hydrophilic polyelectrolytes), the single chain is described as an extended rod-like configuration of electrostatic blobs. [1-3] When water is not a good solvent (hydrophobic polyelectrolytes), the single chain is described in the framework of the pearl-necklace model [4-6]; it is considered as made of a unique type of segments, and the balance between collapse and extension results in the formation of compact beads (the pearls) joined by narrow elongated strings. Simulations [7-11] are consistent with this description of the single chain. Among first reported scattering measurements, some concerned only the upper limit of the dilute regime (close to the semi-dilute regime); they studied partially sulfonated polystyrene, which is a random copolymer of ionisable segments (sulfonated) and hydrophobic segments (not sulfonated). The wormlike chains collapse into more compact objects[12-15], further from each other, which were shown to be made of several pearls[16]. It was also proposed that once the hydrophobic "pearls" formed, the low internal dielectric constant $\varepsilon$[17] may trigger the formation of ion pairs, explaining the observed reduction in osmotically active counterions.[18] Definitely into the dilute regime, two other systems were soon explored, also using SANS and SAXS: they showed agreement with the pearl necklace model. One was obtained by **decreasing solvent quality** through addition to water of miscible bad solvent (or marginal), like acetone: the polyelectrolyte chain undergoes a coil to globule collapse transition [19,20]. Another system in the dilute regime was slightly different: it dealt with the impact of specifically interacting alkaline earth cations which neutralise anionic chains via complex bond formation with the anionic residuals. [21,22]

Other measurements than scattering do not concern dilute solutions but polymers deposited from a solution onto a surface: ellipsometry [13] permits to access to a thickness which depends on rate of charge f, while pearl-like objects are visible by AFM, on systems such as poly(2-vinylpyrridine) and poly(methacryloyloxyethyl dimethylbenzylammonium chloride) [23] and polyvinylamine. [24] More recently an AFM investigation in presence of solvent under different conditions of controlled adsorption (mica as well as lipid membranes) has been conducted on partially sulfonated polystyrene (PSSNa).[25]

Let us turn to solutions in the **semi-dilute regime**, where many chains interact with each other: the highly charged polyelectrolytes in good solvent are described by the isotropic model in which the entangled chains form an isotropic transient network.[1,2] The polyelectrolyte chain is a random walk of correlation blobs, each one is an extended configuration of electrostatic blobs. In **good solvent**, the correlation length $\xi$ i.e. the mesh size of the isotropic transient network scales with polyelectrolyte concentration as $c_p^{-1/2}$ .[1,2] Experimentally, the evolution of $\xi$ with $c_p$ was early verified for highly charged hydrophilic polyelectrolyte, by SANS [26-28], in agreement with the above theoretical prediction. SAXS confirms this [15]; the case of variable charge rates in good solvent regime, controlled via chemical synthesis or variable pH, has been also studied (see for example [29]). The scaling [2] has been extended to account for the partial charge and the Oosawa-Manning condensation [30,31], which agrees with SANS measurements. [32] In the case of **hydrophobic** polyelectrolyte two semi-dilute concentration regimes have been predicted by Dobrynin et al.[6a,6b] : the string controlled regime and the bead-controlled regime. The string controlled regime is similar to the one observed in the dilute regime case; it exists as long as the correlation length $\xi$ of the solution is larger than the string length $l_{str}$ between two neighbouring beads, so a classical polyelectrolyte behaviour is expected, and $\xi$ scales as $c_p^{-1/2}$. For higher concentration where $\xi$ has decreased enough to become of the order of the $l_{str}$, a bead-controlled semi-dilute regime is expected to take place; the correlation length $\xi$ scales as $c_p^{-1/3}$ [6a,6b] and the system behaves as a solution of charged beads of constant size.

This theory has received partial corroboration with the experiments on partially sulfonated polystyrene in water, in semi-dilute regime this time, either by studying the total structure function $S_T(q)$, [14] or the form factor $S_1(q)$. [16] While in dilute solution $S_1(q)$ can be extracted from $S_T(q)$ in some cases, $S_1(q)$ is screened when approaching semi-dilute solution. From the **total structure function** measured by Small Angle X-ray Scattering (SAXS) and also by Small Angle Neutron Scattering (SANS), it was found that $\xi$ scales as $c_p^{-\alpha}$, where $\alpha$ decreases from 0.5 to less than 0.4 when decreasing the chemical charge fraction *f*. [14,16] Later, this was also observed by Baigl et al. through SAXS and Atomic Force Microscopy studies. [33,34] Finally, the counterion condensation in aqueous hydrophobic polyelectrolyte solutions has been studied experimentally through $\xi$, and it was found that the effective charge is strongly reduced [18], compared to the hydrophilic case [30,31] as explained recently.[35] A further step was achieved by **measuring** directly **$S_1(q)$** for partially hydrophobic polystyrene sulfonate-co-styrene: using Small Angle Neutron Scattering combined the Zero Average Contrast method, (ZAC), the interchain correlations could be cancelled, giving interchain correlations, i. e. the form factor, hence the chain conformation. Comparison with totally sulfonated polystyrene (which behaves as a hydrophilic polyelectrolyte in water and shows a wormlike chain conformation [41]) showed with no doubt the rising of a composite strings and pearls conformation [16], in good agreement with the Dobrynin pearl-necklace model and recent simulations which include prediction of the scattering. [11] This enables measuring the size of the pearls, which is of order of several nm and varies with the degree of sulfonation.

Concerning the effect of the solvent quality – i.e. of the level of attractive contribution, we have seen above (for dilute solutions) that it can be **decreased** using solvent mixtures of variable composition, like hydrophilic polyelectrolyte in water plus acetone. [19,20] Conversely, for an hydrophobic polyelectrolyte, the effect of **increasing** solvent quality was investigated experimentally either (i) by increasing the temperature as done by Boué et al [36] on fully sulfonated PSSNa semi-dilute aqueous solutions, showing no effect of temperature on both inter and intrachain interactions [36] or (ii) by using a polar organic solvent of good quality for the backbone allowing at the same time significant ionic dissociation. [37] Case (ii) results in a classical polyelectrolyte-like behaviour, as shown for polyelectrolyte in semi-dilute non aqueous solvent [38] and it was found that the correlation length $\xi$ scales as $c_p^{-1/2}$ as expected for good solvents. [6] On the contrary, in polar organic solvent of poor quality for the polyelectrolyte backbone, two regimes with two scaling exponents $\alpha$ being -1/2 and -1/7 have been observed [38], in qualitative agreement with the predictions of the pearl necklace model. [6] We do not consider here the other possibilities of conformational changes proposed by theories, such as a string – globule coexistence when charges are mobile [39], checked on the corresponding systems [40], but keep in mind, as a starting point, only the pearl necklace-like model, checked formerly on partially sulfonated polystyrene (fixed charges) in water.

In the present work, we investigate the semi-dilute solutions of the same hydrophobic partially sulfonated polystyrene, but in a solvent of improved quality with respect to water, via two routes:

- on the one hand, by adding to water a low proportion of an organic good solvent (for the polyelectrolyte backbone), miscible with water. We vary the charge rates but stay always above the Manning-condensation threshold. The behaviour of PSS will be compared to that of poly(acrylamide-co-sodium-2-acrylamido-2-methylpropane sulfonate), poly-AM-co-AMPS, which is a completely hydrophilic polyelectrolyte in good solvent (water) at the same intermediate degree of sulfonation. In this case we report X-ray Small Angle Scattering, (SAXS), and determine the structure function only.

- on the other hand, the solvent quality is improved by using - instead of $H_2O$ – an other **polar** solvent which is a good organic solvent of the backbone, DMSO. In this case we report SANS, and determine the structure function, and also SANS combined with deuteration and determine the form factor of the polyelectrolyte chain (hence its conformation) in the semi-dilute regime, as a function of polyelectrolyte concentration and the charge rate, by the technique of the Zero Average Contrast (ZAC).

## II-MATERIAL

II-1-Polymer synthesis and characterization

The hydrophobic polyelectrolyte used in this study is a copolymer of styrene and styrene sulfonate (PSS) (poly-(styrene sulfonate)$_f$ – (styrene)$_{1-f}$) with sodium counterions whose chemical structure is shown on figure 1. It was prepared by post-sulfonation of polystyrene based on the Makowski procedure [42, 14], which enables partial sulfonation and leads to a well-defined polyelectrolyte. [43] The Vink method [44] has been also used to synthesize the fully sulfonated PSS ($f = 1$).

The Makowski procedure [42] is a phase transfer, interfacial, reaction. A dichloroethane polystyrene solution is mixed with acetic acid and sulfuric acid in proportions depending on the desired rate of charge. A white layer appears between the two media. After 30 to 60 min, the aqueous phase is neutralized with sodium hydroxide, dialyzed against deionized water until the conductivity of the external dialysis bath remains stable. The solutions are then concentrated with a rotating evaporator and finally freeze-dried. The resulting white powder is better stored away from light.

The Vink method [44] is commonly used to reach total sulfonation; one starts from a polystyrene (PS) solution in cyclohexane (a PS theta solvent at 35°C), which is poured onto a mixture of sulfuric acid with phosphoric acid. After stirring for half an hour, the mixture is let to rest for decantation. Separation in three phases is triggered by addition of ice. The phase containing PSSH (polysulfonic acid) is extracted, neutralized by an excess of sodium hydroxide. The obtained PSSNa solution is dialyzed, concentrated and freeze dried.

The ionisable unit (SS), or sulfonation rate $f$ of the polyelectrolytes was varied between 0.3 (the limit for solubility in water) and 1 (fully charged). The sulfonation rate $f$ is thus always above the Manning condensation limit for the charge rate, equal to $a/l_B \sim 0.3$ for PSS in water (a length of one unit, $l_B \sim 7$ Å, Bjerrum length in water). Note that $l_B$ is larger in DMSO (12 Å), and that we will find a lower value for a (see section IV), so that the condensation threshold should be lower.

For the zero average contrast technique to apply correctly, the mass distribution should be very close for the perdeuteriated polystyrene (d-PS) and the non-deuterated polystyrene (h-PS), and the masses must be well defined to allow data fitting. Polymers with very close degree of polymerization (d-PS, $N_{wD}$ =652, h-PS, $N_{wH}$ = 625) and a narrow mass distribution about 1.03, were purchased from Polymer Standard Service (Mainz, Germany). We also checked that the final sulfonation rates are very close for the deuterated and the non-deuterated chains (within 1 or 2 %, see Table1). We preferred to check d and h polymer separately because sulfonation can lead to different result depending on the initial degree of aggregation, and to control any slight isotopic effect. In practice after sulfonation of d-PS, several trials were conducted on h-PS until obtaining the same values of $f$.

For X-rays, the polystyrene used had a higher mass and a broader mass distribution ($M_w$ = 250 000, with a polydispersity of 2).

The hydrophilic polyelectrolyte also studied in this paper is poly- co-(sodium-2-acrylamido-2-methylpropane sulfonate)$_f$ -(acrylamide).$_{(1-f)}$, whose chemical structure is shown on figure 1. The average molar mass of the monomer is 71 +158.$f$. It was synthesized by radical copolymerisation of acrylamide with acrylamido-2-methyl propane sulfonic acid [45], which was slightly modified by adjusting the ratio of the two monomers to obtain, after neutralisation, a fraction of ionisable unit (AMPS),$f$, between 0.3 and 1. Thus these chains are also highly charged polyelectrolytes. The resulting molecular weight is $M_w = 650,000$ ($N_w$ 2800) and the polydispersity is 2.6, for $f = 1$. Note that both polyelectrolytes (AMAMPS and PSS) are salts of strong acid, bearing $SO_3^-$ anions as side groups when ionised, with $Na^+$ counterions. So the two polyelectrolytes used in this study differ mainly by the solvation characteristics of their non ionisable units (hydrophobic in the case of PSS, hydrophilic in the case of AMAMPS).

The corresponding characteristics of all these polymers are summarised in table 1.

II-2. Preparation of solutions

Solutions were prepared by dissolving dry polyelectrolyte (assuming slightly more than 10% of residual water in the powder, as suggested recurrently through various techniques and confirmed as 10 to 14% by Karl Fisher tests at the Laboratoire Central d'Analyse LCA/USR59/CNRS Vernaison) – in the solvent and letting at rest for two days before measurements. For SAXS measurements, the solvent is deionised $H_2O$, or a mixture of deionised $H_2O$ and tetrahydrofuran (THF). For the SANS measurements, the solvent is dimethylsulfoxide DMSO, either in non deuterated form, h-DMSO ($SO(CH_3)_2$), or in deuterated form, d-DMSO ($SO(CD_3)_2$; (see table 2 for its characteristics) or in their mixture with suitable composition for the zero average contrast method. It is interesting to note that dissolution was immediate for deuterated PSS $f = 0.64$ in both normal h-DMSO, and non deuterated PSS in pure d-DMSO, whereas for $f = 1$ and 0.36 it took a few hours for deuterated PSS in normal h-DMSO, and two weeks for normal PSS in d-DMSO.

The question of whether the solutions concentrations are below (dilute) or above the overlapping concentration (semi-dilute) cannot be answered completely at this stage. We can only recall the former results in water. For AMPS, which is completely hydrophilic, the solutions without salt are always semi-dilute: the number of units is ~ 3000. To give an order of magnitude, with an electrostatic blob of 60 units assuming $f$ at its lowest value 0.3, and a blob size = 20 Å, the fully extended length is ~ 150 x 20 = 1000 Å, which gives $c^* \sim M/(4\pi.L^3/3) = 635000/(6.10^{23} \cdot 4\pi.(10^9/8) 10^{-24})/3 = 2 \cdot 10^{-3}$ g/cm$^3$. For PSS, we have calculated in Ref. (16) the radius of the part of space available per chain, $R_{overlap}$, as a function of $c_p$ (89 Å for $c_p = 0.34$ M and 112 Å for 0.17M). If $R_g > R_{overlap}$, the solution is semidilute. We concluded that solution was semidilute for f = 0.64 and 0M salt but slightly dilute in the other case (for $c_p = 0.34$ M, with 0 M added salt, and for both $c_p$, with added salt concentration $c_s = 0.34$ M or more). The regime obtained in this paper will depend on the value of $R_g$. In DMSO this will be measured and we will see that solutions are always semi dilute. In THF + water mixtures, without salt, we will also see that the chain increase in size, so that for the concentration used here, $c_p$ = 0.32 M, 0 M salt, the solution is likely to pass in the semidilute regime.

For SANS, all solutions in DMSO were filtrated, except the one for $f$ = 0.36 at $c_p$ = 0.34 M, for which it was impossible. Such filtration was shown to abate noticeably the low q upturns in water at any $f$. Conversely, for $f$ = 0.64 in DMSO, no effect was observed for the scattering.

All concentrations are expressed in M/L.

## III-METHODS

### III-1-SAXS measurements

Small Angle X-ray Scattering (SAXS) experiments were performed on beam line D22, at LURE, using the DCI synchrotron radiation source. Data were obtained with pinhole collimation and recorded with a linear detector of 512 cells. The scattering vector $q$ varied from 0.008 to 0.2 Å$^{-1}$ [$q = (4\pi/\lambda \sin(\theta/2))$, where $\theta$ is the observation angle and $\lambda$, the wavelength was 1.37 Å]. The scattering data were normalized to constant beam intensity and corrected for transmission, sample thickness, parasitic and background scattering. The resulting scattering profiles are obtained as normalized intensities in relative units versus scattering vector $q$.

### III-2-SANS measurements

**III-2.1 The Zero average contrast method:**

Let us recall the fundamentals of the most convenient method by which the form factor of a chain among others can be obtained. We start from the general expression of the scattered intensity:

$$I(q) \text{ (cm}^{-1}, \text{ or Å}^{-1}) = (1/V) \cdot d\Sigma/d\omega = I(\vec{q}) = \frac{1}{V}\left\langle \sum_{i,j} k_i k_j \exp(i\vec{q}(\vec{r}_i - \vec{r}_j)) \right\rangle \quad (1a)$$

Here $k_i$ (cm or Å) = $b_i - b_s (V_{mol\,i}/V_{mol\,s})$ is the "contrast length" between one repeating unit of scattering length $b_i$ and molar volume $V_{mol\,i}$, and a solvent molecule ($b_s$, $V_{mol\,s}$).

Consider first the case where all chains are labelled with respect to solvent; in practice we dissolve H-PS into perdeuterated D-DMSO (this case applies also for X-rays with H-PS in water or water + THF). The concentration is $c_p$, in mole/L (or mole/Å$^3$), so the total volume fraction of chains is $\Phi_T = N_{Av} \cdot c_p \cdot V_{mol\,i}$, where $N_{Av}$ is the Avogadro number. Then for all i, we have $k_i = k_H$ (the values are given in Table 3), and

$$I(q) \text{ (cm}^{-1}\text{, or Å}^{-1}\text{)} = (1/V) \cdot d\Sigma/d\omega = k_H^2 \, S_T(q) \quad (1b).$$

Using Å and Å$^{-1}$ as the units for $k_H$ and $I(q)$, we obtain $S_T(q)$ in Å$^{-3}$. Quite generally,

$$S_T(q) = S_1(q) + S_2(q), \quad \text{Å}^{-3} \quad (2a),$$

where

$$S_1(q) \text{ (Å}^{-3}\text{)} = \frac{1}{V} \left\langle \sum_{\substack{\alpha \text{ avec} \\ \beta = \alpha}} \sum_{i,j} \exp(i\vec{q}(\vec{r}_i^{\,\alpha} - \vec{r}_j^{\,\beta})) \right\rangle \quad (2b)$$

corresponds to the correlations between monomers i,j of the same chain $\alpha = \beta$ (intrachain scattering) and

$$S_2(q) \text{ (Å}^{-3}\text{)} = \frac{1}{V} \left\langle \sum_{\substack{\alpha, \\ \beta \neq \alpha}} \sum_{i,j} \exp(i\vec{q}(\vec{r}_i^{\,\alpha} - \vec{r}_j^{\,\beta})) \right\rangle \quad (2c)$$

corresponds to the correlations between monomers i,j of two different chains $\alpha$ and $\beta \neq \alpha$ (interchain scattering).

Consider now the case where *only a fraction of the chains are labeled*. We use a mixture of a number fraction $x_D$ of d-PSS chains ($k_i = k_D$) and $(1-x_D)$ of h-PSS chains ($k_i = k_H$). The total volume fraction of chains in the solution is the sum of the volume fractions of the two types of chain, $\Phi_T = \Phi_H + \Phi_D$ (we have in general $V_{molH} = V_{molD}$, so $\Phi_D/\Phi_T = x_D$ and the equation $\Phi_T = N_{Av} \cdot c_p \cdot V_{mol\,H}$ is still valid, $c_p$ being the total polymer molar concentration). The scattered intensity (1a) becomes:

$$I(q) \text{ (cm}^{-1}\text{)} = (1/V) \cdot d\Sigma/d\omega = \{[(1-x_D)\,k_H^2 + x_D\,k_D^2]\,S_1(q)\} + \{[(1-x_D)\,k_H + x_D\,k_D]^2\,S_2(q)\}$$

(3a)

This second type of labeling allows us to suppress the interchain contribution $S_2(q)$, if we can have

$$(1-x_D)\,k_H + x_D\,k_D = 0. \quad (3b)$$

This is possible if we use as a solvent a mixture of H-DMSO / D-DMSO : then the average scattering length of the solvent $b_S$ can be varied. In the equation above, the symmetric case $k_H = -k_D$ (which also implies $x_D = 0.5$) is the most efficient situation in term of intensity. This is obtained if $b_S/V_S$ is made equal to the arithmetic average of $b_H/V_{mol\,H}$ and its pendent $b_D/V_{mol\,D}$. For h-PSS and d-PSS, this corresponds to a solvent constituted of 19% H-DMSO and 81% D-DMSO. We write $|k_{ZAC}| = -k_H = k_D$, and Eq. (3) gives :

$$I(q) = k_{ZAC}^2 \, S_1(q) \qquad (4)$$

which permits a direct measurement of intrachain scattering of one chain among the others, even in the semi-dilute regime. The different contrast length values are listed in Table 2. The values evaluated for the contrast lengths of the Na counterions with the DMSO-H/ DMSO-D mixture used here are low; their contribution to the scattering have therefore been neglected. This has been confirmed by a more refined evaluation accounting for hydration [46,47]. The $S_1(q)$ limit at q tending to zero is :

$$\lim_{q \to 0} S_1(q) \, (\text{Å}^{-3}) = c_p \, (\text{mole/Å}^3) \, N_{Av} \, N_w \qquad (5)$$

where $c_p$ should be expressed in mole/Å$^3$ = $10^{-24}$ (M/L). Hence, from the definition of the form factor, we can write

$$S_1(q) = c_p \, N_{Av} \, N_w \, P(q) \qquad (6)$$

The ZAC technique has been used since on polyelectrolytes by other authors. [47-49]

**III-2.2 Measurements and data treatment:**

SANS measurements have been performed on the PACE spectrometer at the Orphée reactor of LLB, CEA- Saclay, France (www-llb.cea.fr). A range of scattering vector $q = (4\pi/\lambda) \sin\theta/2$ between $5.10^{-3}$ and 0.4 Å$^{-1}$ was covered using the following two settings: D=0.92m, $\lambda$=5Å and D = 3.02m, $\lambda$=12.5 Å. Samples were contained in 2 mm thick quartz cells. All measurements were done at room temperature.

All data have been normalized using the incoherent scattering of a high proton content sample, here light water; the latter has been calibrated to obtain absolute values of $(1/V).d\Sigma/d\omega_{water}$ in cm$^{-1}$, using Cotton's method. [50] The solvent contribution is subtracted from these corrected data. The subtraction of the solvent incoherent background is however quite delicate and deserves further remarks. At large q (> 0.2 Å$^{-1}$) especially, the coherent part of the intensity is very small compared to the background due to incoherent scattering, which has several origins:

- incoherent scattering from $H_2O$ and $D_2O$ in the solvent,

- hydration water molecules adsorbed on the polymer dry chains (i.e. "residual water", more than 10% in weight).

- incoherent scattering from protons in the h-PSSNa, and deuterons in the d-PSSNa.

- protons from water vapor molecules after contact with air.

These small contributions are delicate to estimate and thus make us unable to know the exact quantity to subtract with accuracy better than 3%.

Such uncertainty has little influence for small q but can lead to different shapes of the scattering curves for large q.

Also, mixing the components leads to an extra flat scattering, called Laue scattering or sometimes "mixing incoherent", which is actually the scattering from the mixture of small elementary components such as different molecules in a solvent.

For best results, and to eliminate as much as possible effects of multiple scattering (though they are here very weak) which involve the cell geometry, we have prepared under the same conditions some blanks, by mixing a non deuterated and a deuterated solvent, D-DMSO with H-DMSO aiming at the same flat intensity, and therefore the same neutron transmission. This is particularly sensitive for the measurements of $S_T(q)$, where the solvent is deuterated and the polymer, non-deuterated, brings an important incoherent contribution. We took the incoherent in consideration, to make our blanks, by introducing the same amount of protons through a volumic fraction $\phi_H$ of non deuterated solvent: the value is equal to $\phi_H = 0.01$ for the concentration $c_p = 0.34$ M/L, and to $\phi_H = 0.02$ for $c_p = 0.5$ M/L.

## IV-RESULTS AND DISCUSSION

### IV-1-Polyelectrolytes in water plus organic solvent, THF:

The organic solvent selected to be added in small proportions to water is tetrahydrofuran (THF). It is partly polar since its dielectric constant ε is 7.6, whereas it dissolves the backbone i.e. pure polystyrene (non sulfonated) as well as the partially sulfonated PSS for rates $f \leq 0.6$ at a polymer concentration of 0.32 M/L. However, pure THF does not dissolve the hydrophilic polyelectrolyte AMAMPS for all the charge rates at the same polymer concentration of 0.32 M/L.

Note that the addition of THF in $H_2O$ is done in small or moderate proportion, so that the PSS or the AMAMPS behave as a polyelectrolyte; in particular it has been checked by viscosimetry that dilute polyelectrolyte solutions exhibit the classical behaviour of polyelectrolytes, close to the one of other charged colloids: the reduced viscosity increases as the polyelectrolyte concentration decreases either in $H_2O$ or in the mixture of $H_2O$ and THF.

We note $\phi_{THF}$ or "% THF" the percentage by volume of added THF.

#### IV-1-a- Effect on a hydrophilic polyelectrolyte for different charge content

The SAXS profiles showing the effect on structure of addition of THF on the structure for the hydrophilic polyelectrolyte AMAMPS, for different charge rates are shown in figure 2. The polyelectrolyte concentration is kept equal to 0.32 M/L, for all solvent compositions.

It emerges that *the scattering is independent of the addition of THF. The curves for 0% and 25% THF just overlap.* Thus the peak position does not depend on the amount of added THF, which shows that the polyelectrolyte chain network of the AMAMPS remains unchanged. Moreover the constancy of the peak intensity suggests that the effective charge and the contrast chain/solvent remain constant with THF addition (for the SAXS data discussed here, the contrast comes from the condensed metallic counter-ions). The constancy of the width of the peak confirms that the order degree of the system remains constant. In this frame, the fact that the scattered intensity at zero angle is constant with the addition of THF, also supports the idea that the effective charge is constant: the scattered intensity at zero angle is related to the osmotic compressibility as, according the Dobrynin model[2] :

$$S(q \to 0) \approx kTc_p \frac{\partial c_p}{\partial \Pi} \approx c_p / f_{eff} \qquad (7)$$

The absence of low q upturn signals the absence of large hydrophobic aggregates, which are likely to dissolve in the presence of organic solvent.

What must be retained is that the chain network structure is independent of moderate addition of THF for the hydrophilic polyelectrolyte.

### IV-1-b- Effect on a hydrophobic polyelectrolyte.

Figure 3 shows the SAXS profiles as a function of THF addition in the hydrophobic PSS solutions, for various degrees of sulfonation. Figure 3a shows that for the fully charged PSS ($f = 1$), the position, the width and the intensity of the peak remain unchanged when adding 12.5 and 25% of THF. However, the scattered intensity varies at lower q: the depletion in the curve, before the upturn at $q \to 0$, is more pronounced with increasing quantity of THF.

On the contrary, when $f < 1$ (for $f = 0.58$ and even more for $f = 0.38$, Fig. 3b and c), the structure strongly varies with addition of THF. Obviously, the peak position q* increases and its height decreases. Also, the peak widens: when passing from 0% to 25% THF, $\Delta q/q^*$ increases from 0.55 to 0.85, for $f = 0.58$ and from 0.67 to 0.72 for $f = 0.38$ (q* and $\Delta q$ are determined within uncertainty due to the low q upturn). We pass from repulsion, like between two charged spheres, to a softer effect like for a network of interpenetrated charged chains. In the low q region, the depletion of the scattered intensity is deeper, suggesting that the system becomes less compressible, also like when chains are more charged. The upturn at $q \to 0$ is also reduced. This suggests a decrease of the large scale inhomogeneities in the solution.

This evolution of the profiles with adding THF is similar to the one in pure water with increasing the charge rate *f*. Values for a given *f* at 25% THF rejoin values at the next highest *f* at 0% THF. For example, the profile for *f* =0.38 and 25% THF is close to the one for *f* =0.58 in water. This effect can be seen quantitatively at a glance in Figure 4 showing the shift of the peak abscissa $q^*$, the decrease of the peak intensity $I(q^*)$ and the decrease of $I(q \rightarrow 0)$. For f = 0.58, $q^*$ and $I(q^*)$ for 25% THF are very close to the values for f=1 in water. We know from SANS that in this case of *f* = 1 in water the chain conformation is string-like [41].

At large q > 0.1 Å$^{-1}$, all the scattering profiles coincide indicating that at small spatial scales the monomer-monomer correlations are independent of THF addition. In summary, the addition of THF causes an evolution of the structural characteristics towards the ones of a more charged polyelectrolyte in water. This evolution suggests a decrease of the pearls.

Within the framework of the Dobrynin model[2] describing the structure of hydrophobic polyelectrolyte in poor solvent as PSS in H$_2$O, the maximum in the scattering intensity profile scales as (Eq.(15) and (8) of Ref. (2)) :

$$q^* \propto \left( \frac{B}{c_p a} \right)^{-1/2} \propto f_{eff}^{2/3} \left( \frac{c_p a}{\tau} \right)^{1/2} \varepsilon^{-1/3} \qquad (8)$$

in the regime T lower than the Θ temperature ( $\tau = (\Theta-T)/\Theta$ )).

At first thought, we would imagine adding THF would decrease the dielectric constant ε, therefore would decrease the effective charge $f_{eff}$ and thus increase dipole-dipole interactions. This would reduce the solvent quality (increase in τ), knowing of course that what happened in reality is the opposite. Since it is difficult to estimate the respective weight of each quantity, we cannot predict simply the evolution of $q^*$ according to Eq.(8).

However, the reality is different and more simple: taking the case of AMAMPS or the case of fully sulfonated PSS, results deny any direct effect of the variation of these three parameters with addition of THF. Concerning the dielectric constant ε, this is expectable: the weak dependence of $q^*$ with ε ($\varepsilon^{-1/3}$) cannot lead to a noticeable change for such small amounts of THF. This variation would be even smaller if the effect of THF addition is local (like a sorption effect, see below), so that from a global view, the dielectric constant remains equal to that of H$_2$O. Concerning $f_{eff}$ and τ, no effect of reduction of polarity or of solvent quality is seen: on the contrary, the decrease of the scattered intensity at low q with THF addition, for *f* =1, is a sign of better solvent quality (less aggregation at large scale), while electrostatic repulsions remain identical (same value of $f_{eff}$ according to Equation 7, and in agreement with the fact that the peak shape is unchanged).

Only for PSS $f < 1$ is seen a difference in behaviour under THF addition. This shows that THF acts on the hydrophobic regions. This prompts us to explain the results by another process: when THF is added to aqueous solutions, the polyelectrolyte tends to behave like a classical polyelectrolyte, because the solvent is better. Acting most probably on the hydrophobic domains in a local way, the THF solubilises these domains. Because this results in fewer pearls and more strings, the chain conformation is more stretched. As a consequence we observe a return towards the characteristics of fully sulfonated PSS, therefore a decrease of q* ($\tau$ is reduced and we leave the regime T < Θ). At the same time, the counterions, initially localised in $SO_3^-/Na^+$ ion pairs inside or condensed at the surface of the hydrophobic domains, become surrounded by solvent. Since this solvent is mostly an aqueous medium, these ions pairs dissociate or "decondense"; such release gives rise to an increase of the effective charge of the chain. This is an indirect effect of THF (although it is less polar *per se*).

At this stage, we have to consider a possible preferential sorption of one of the component of the solvent composition. Namely a fraction of THF could adsorb preferentially on the hydrophobic domains of the polyelectrolyte chain, which would cause their dissolution. This would imply that the polyelectrolyte is globally surrounded by a solvent richer in water. All together this does not modify the resulting chain extension. From the point of view of scattering technique, preferential sorption of THF (or water) could vary the contrast between chain and solvent. Since THF and $H_2O$ have similar scattering densities, the effect should not be strong. In practice we see no effect for $f = 1$.

But we can even make further conclusions if we compare with DMSO (next Section). It will be shown that using a solvent made of a single chemical species leads to very similar effect. This does not support sorption in the case of THF.

In summary, for the hydrophobic polyelectrolyte, the solvent quality is improved when adding THF and the polyelectrolyte tends to a structure characteristic of more charged and elongated chains, like in water for non hydrophobic polyelectrolyte ($f = 1$).

IV-2- PSS in both polar and organic solvent, DMSO.

We are interested in investigating the behaviour of PSS hydrophobic polyelectrolyte in a pure solvent better than water, i.e. in a media where the affinity of every solvent molecule is increased for the polyelectrolyte backbone and especially for the non charged monomers. The effect of the charge rate (above the Manning condensation threshold) and polyelectrolyte concentration can thus be studied without the side effects brought by bad solvent. The organic solvent should be polar to dissolve the electrostatic charges, but also aprotic to avoid the hydrophobic effects observed in solvents which are structured by hydrogen bonds. [51] The chosen polar aprotic solvent is dimethylsulfoxide (DMSO): its dielectric constant ε equals 46, it dissolves easily the partially sulfonated polystyrene at the three charge densities studied, $f$ =0.36, $f$ =0.64 and $f$ =1 (although it does not dissolve the neutral polystyrene).

In this study we have measured both the *total scattering* –measured here using neutron radiation, instead of X rays, and the *single chain scattering*, which can be reached only using neutrons.

### IV-2-a-Effect of polyelectrolyte concentration for the chemical charge $f$ =0.64:

*-Total scattering* :

For the charge content $f = 0.64$, the hydrogenated polymer concentration in the deuterated DMSO was varied in a range between 0.085 and 0.5 M/L. The set of the scattering profiles $S_T(q)/c_p$ is shown on Figure 5. For all polyelectrolyte concentration, a maximum appears in the scattering profile at a finite scattering vector q* whose position shifts to higher q as $c_p$ increases. The position of the peak q* is found to vary as $c_p^{0.44}$ which is not far from the classical evolution of q* ~ $c_p^{0.5}$ usually found for highly charged polyelectrolytes. Moreover, the scattered intensity per monomer $S_T(q)/c_p$ decreases as $c_p$ increases and the peak broadens. For q>q*, all the scattering profiles $S_T(q)/c_p$ coincide for all $c_p$: at small spatial scales the monomer-monomer correlations for $S_T(q)/c_p$ are independent of $c_p$ which indicates a good dissolution of the polyelectrolyte in DMSO. The minimum of the normalised scattered intensity is constant. If we assimilate it to the theoretical limit at q=0, $S_T(q \rightarrow 0)/c_p$, this constancy is in agreement with the Manning condensation theory, following which it should be equal to $kT/f_{eff}$.[31] For the highest concentrations, the precise determination of $S_T(q\rightarrow 0)/c_p$ is prevented by the upturns observed at very small q (q→0). However we notice in passing that these upturns are clearly lower than for the fully charged polyelectrolyte (f=1) in $H_2O$ indicating that there is less aggregation in DMSO than in $H_2O$.

Apart from this last detail, in summary, the total scattering for f = 0.64 in DMSO is similar to the one of $f = 1$ in water, as can be seen by comparison with former SANS results, like Fig. 1 of Ref.[16] (same q* ~ 0.75 Å$^{-1}$ for $c_p = 0.34$ M, comparable height after division by $c_p$).

*-Zero Average Contrast (ZAC) conditions: the Chain form factor from $S_1(q)$*

Figure 7 shows in log-log plot the scattering profiles $S_1(q)/c_p$ as a function of polymer concentration, for the PSSNa f= 0.64 under ZAC conditions in a log-log representation. All the scattering profiles are superimposed at large q in the asymptotic domain and $S_1(q)/c_p \propto q^{-1}$ (rodlike conformation). At small angles q→0, the scattering profiles converge, as required, to the same value $N_w$ of the degree of polymerisation according to Eq. 5: $S_1(q\to 0) = c_p N_{av} N_w$. It was found that $N_w = 720$, which is the same value in $H_2O$ for that polymer [16], as required also. This is a proof that we measure the single chain scattering.

The form factor of the chain determined experimentally according to Eq 6 was fitted according to Sharp and Bloomfield (SB) model [52] describing the polyelectrolyte chain as wormlike with a finite length L and a persistence length $l_p$. The SB form factor is given by :

$$P(q) = \frac{2[\exp(-x) + x - 1]}{x^2} + \left[\frac{4}{15} + \frac{7}{15x} - \left(\frac{11}{15} + \frac{7}{15x}\right)\exp(-x)\right]\frac{2l_p}{L} \quad \text{with } x = \frac{Lq^2 l_p}{3} \quad \text{Eq.9}$$

Figure 7 shows how well the SB model fits the obtained results, in the $q^2 S_1(q)$ vs q representation (we chose the examples of the two extreme polymer concentrations, $c_p = 0.085$ M $l^{-1}$ and $c_p = 0.85$ M $l^{-1}$). Moreover, using the universal *des Cloizeaux* representation [53] $q^2 L l_p P(q)$ as a function of $q.l_p$, in Figure 8, we see that all the curves are superimposed at large q. This is exactly what has been observed in water for the fully charged polyelectrolyte PSSNa $f = 1$. [41] This confirms the consistency of the extracted values of the persistence length. Plotting these values of $l_p$, as a function of the ionic strength $I_S$ on Figure 9 ($I_S = fc_p + 2c_s$, where $c_s$ is the concentration of the external added salt and here $c_s = 0$), we find that $l_p \sim (fc_p + 2c_s)^{-0.41}$. This is not very far from the variation of $l_p$ for $f = 1$ found in water ($l_p \sim (fc_p + 2c_s)^{-1/3}$ in water, if –like here- we do not subtract the intrinsic persistence length $l_0$). [41]

In addition to the fit to the SB model, the radius of gyration can be directly determined from the Zimm plot in the Guinier domain as follows:

$$\frac{1}{S_1(q)} = \frac{1}{S_1(0)} * (1 + q^2 R_G^2 / 3) \quad \text{for } q.R_g < 1 \quad \text{Eq. 10}$$

The radius of gyration scales as $(f.c_p + 2c_s)^{-0.23}$ (Figure 9), so $R_g$ is proportional to $l_p^{1/2}$ which is in agreement with the behaviour of a wormlike chain.

From the fits, it emerges that the total length of the chain, L, is smaller in DMSO than in $H_2O$; in agreement with this fact, $R_g$ is smaller too. This is consistent with the fact that the interchain distance is higher as signalled by the scattering maximum abscissa q* value which is lower in DMSO. The length per monomer $a = L/N_w$ decreases from 2 Å in $H_2O$ [16] to 1.6 Å in DMSO; this value is the same than for PSS in water with tetramethyl ammonium as counterions.[54] So the linear density should increases in DMSO, which is also seen in the asymptotic domain ($ql_p>1$): we see in the log-log plots of Figure 10 a shift between $S_1(q)_{DMSO}$ and $S_1(q)_{H2O}$ corresponding to a factor $S_1(q)_{DMSO} / S_1(q)_{H2O} \approx 1.26$ which corresponds to the ratio of $a_{H2O}/a_{DMSO}$. The same value of $a$ was found for PSSNa in water with TMA counterions and attributed to a helical structure.[54] This suggests that DMSO can structure locally the chain.

Finally, an additional quantity obtained from scattering is the apparent structure function which we assume to be given by $S(q) = S_T(q)/S_1(q)$, where $S_T(q)$ is the total scattering function and $S_1(q)$ is the intrachain scattering function. Figure 11 shows the evolution of the structure function $S(q)$ as a function of polyelectrolyte concentration for PSS $f=0.64$ in DMSO. As $c_p$ decreases, the maximum of the signal is more marked and its abscissa decreases, which means that the interchain correlation distance increases. This behaviour is the same as encountered in $H_2O$.[16] As also seen in water, the values of the maximum of the signal in the apparent structure function are usually higher than the maximum of the total signal function. However, these two maxima scale in the same way with the polyelectrolyte concentration $c_p$ (the insert on Figure 11).

In summary, the behaviour for $f=0.64$ in DMSO is very close to the one for $f=1$ in water.

<u>IV-2-b-Effect of the variation of the chemical charge rate of the polyelectrolyte :</u>

*-Total scattering* :

Figure 12a shows the evolution of the scattering profiles as a function of the charge rate for the deuterated polyelectrolyte in hydrogenated DMSO for $c_p = 0.17$ M/L. It is obvious that for $c_p = 0.17$ M/L, all the scattering profiles for the different charge rates are superimposed. This is in perfect agreement with the charge renormalisation law, since the effective charge $f_{eff}$ should be the same for the three values of f after Manning condensation. For higher $c_p = 0.34$ M/L, Figure 12b shows that the scattering profile for $f =0.36$ remains roughly superimposed with that of $f =0.64$. However, the scattering peak for $f =1$ shifts to lower q and the intensity increases.

Figure 12c shows the evolution of the scattering profiles as a function of the charge rate for the reverse isotopic labelling, i.e. non deuterated polyelectrolyte in deuterated DMSO at a concentration $c_p = 0.34$ M/L. Here, the behaviour is different for each *f*. For $f = 0.64$ the position of the peak in the scattering profile remains the same as hydrogenated DMSO. For *f*

=0.36, we see a strong upturn: this is just due to our inability in filtering the solution. For $f$ =0.36 as well as $f$ =1, the maximum of the peak shifts to lower q compared to $f$ = 0.64 and the intensity is higher. This behaviour can be ascribed to an incomplete solubilisation of the polyelectrolytes for both $f$ = 0.36 and $f$ = 1. These discrepancies match exactly the very long times necessary for dissolution for both sulfonation rates, reported in section II. Recall that for a given polyelectrolyte and at a given temperature, the solvent quality can be varied by its deuteriation and the theta temperature should increase from hydrogenated to deuteriated solvent. In the case of PSS, the polyelectrolyte solutions of extreme charge rates are more sensitive to the isotopic nature of DMSO than that of $f$ =0.64, because closer to a theta temperature. The origin of the incomplete solubilisation is different for PSS $f$ =0.36 (high ratio of non charged monomers) and PSS $f$ =1 (high ratio of charged monomers). All these features are attributed to complex effects of solvent structure.

*-Zero Average Contrast conditions: the Chain form factor from $S_1(q)$*

Figure 13 shows the intrachain scattering $S_1(q)$ (proportional to the form factor) at the different charge rates $f$ at a polyelectrolyte concentration $c_p$ = 0.34 M/L. The spectra for all $f$ values are superimposed in a large domain of wave vector q (for q > 0.015 Å$^{-1}$) and the problems of bad solubility and aggregates reported just above for the total scattering function $S_T(q)$ in pure D-DMSO, don't reverberate on the chain form factor in mixtures of H-DMSO and D-DMSO. The conformation of the chain seems to be independent of the charge rate in DMSO. This in perfect agreement with the fact that the effective charge $f_{eff}$ should be the same for the three values of $f$ after Manning condensation.

At small angles (q → 0), the scattering profiles converge to the same value $N_w$ of the degree of polymerisation according to Eq. 5: $S_1(q \to 0) = c_p N_{av} N_w$. In DMSO, it is found that $N_w$ = 640 for PSS $f$ =1 and $N_w$ = 850 for $f$ =0.36. These values of $N_w$ are lower than those obtained in H$_2$O where $N_w$ = 850 for PSS $f$ =1 and $N_w$ = 1130 for PSS $f$ =0.36. The decrease of $N_w$ and so of the apparent $S_1(q \to 0)$ is attributed to a reduction of permanent aggregates in organic solvent compared to H$_2$O.

Concerning the persistence length, we find that in DMSO $l_p \sim I_S^{-0.4}$ (where $I_S$ is the ionic strength of the solution $c_p+2c_s$; as defined above), for all the charge content $f$. As observed above for $f$ = 0.64, this variation is similar to that in H$_2$O where $l_p \sim I_S^{-0.33}$ was found. [37] However, the absolute values of the persistence length in DMSO for PSS $f$ = 0.64 are lower by a factor of 1.5 for PSS $f$ =1 in H$_2$O (Figure 14). If we consider that all the counter ions are dissociated in H$_2$O and DMSO, this factor 1.5 disagrees with the hypothesis that the persistence length is only related to a distance between ions in the solution. In this case, the two curves should be superimposed. So, an additional factor K related to the solvent or to the local structure of the polymer in the solvent – may be in relation with what observed in IV-2-a for L and $a_{DMSO}$ - should intervene and the persistence length can be expressed as $l_p = K I_S^{-0.4}$.

As previously, we assume that the structure function is given by $S(q) = S_T(q)/S_1(q)$ where $S_T(q)$ is the total scattering function and $S_1(q)$ is the intrachain scattering function. Figure 15 shows the evolution of the structure function $S(q)$ as a function of charge density in DMSO for a polyelectrolyte concentration of $c_p=0.34$ M l$^{-1}$. It arises that for all charge rates, the different profiles $S(q)$ are very similar (except for the upturn at low q for $f = 0.36$ (not filtered). The values are very close to 1 at large q, for all samples in DMSO. The profiles are also very similar to the one for a fully charged PSS in H$_2$O i.e. a classical polyelectrolyte) like for $S_T(q)$ and $S_1(q)$. Looking more in detail, we note that the peak of $S(q)$ is less marked in DMSO than in H$_2$O for PSS $f = 1$ without added salt. This difference is not due to a decrease of the maximum of the peak intensity but comes from the large angles region where the decrease of the signal $S(q)$ is slower. This feature can be due to the following effect: the large q scattering for $S_T(q)$ and $S_1(q)$ may be slightly different because the deuterated fractions $x_D$ in the two solvents used are slightly different. Since these two fractions $x_D$ are less different in DMSO (in ZAC solvent $x_{D-DMSO} = 0.81$, and in the solvent giving the total signal $x_{D-DMSOd} = 1$) than in H$_2$O (in ZAC solvent $x_{D2O} = 0.71$ and in the solvent giving the total signal $x_{D2O} = 1$), the ratio $S_T(q)/S_1(q)$ stays closer to 1 in DMSO.

**V-CONCLUSION :**

We have shown in this study that the solvent quality for the hydrophobic polyelectrolytes in aqueous solutions can be improved by adding to water a miscible organic good solvent of the backbone - in low proportion, or by using an organic polar solvent instead of water.

**Concerning the first case**, i.e. the addition of low proportion of water miscible organic solvent to aqueous solvent, we obtained measurements of **$S_T(q)$ only (SAXS)**. They show that the structure of PSS solutions is progressively and significantly modified upon the addition of THF towards the behaviour observed for f = 1 in water. Indeed, the maximum of the "polyelectrolyte peak" is shifted to higher q, the intensity of peak decreases and the scattered intensity at zero angle decreases too. This evolution is more pronounced as the charge rate is lower. We know from former $S_1(q)$ measurements in pure water that partially sulfonated PSS is pearl necklace-like. We do not follow the conformation here, but results on $S_T(q)$ suggest on a sensible basis that, the chain conformation, evolves towards a more stretched conformation as that in pure water for fully sulfonated PSS"

Conversely, for the hydrophilic polyelectrolyte (AMAMPS) aqueous solutions, the behaviour is markedly different since the structure function remains constant upon addition of small amounts of THF, for all charge rates. As discussed in the text, this means that the reduction in dielectric constant upon addition of THF, which would act in disfavor of charge dissociation, is not important. On the contrary, this supports the idea that THF effect on partially sulfonated PSS is linked with its hydrophobicity: the mechanism of decompaction of the chain, can be proposed as involving two steps. First the dissolution of the hydrophobic domains, which in second can make free the counterions condensed earlier on these domains, and promote an increase of the polyion charge and the return to an extended state.

**Concerning the second case,** i.e. partially sulfonated PSS in a non selective solvent, DMSO, we obtained SANS measurements of both **$S_T(q)$ and $S_1(q)$**. What is seen indirectly for THF can be seen directly for DMSO. It erases quasi-completely the hydrophobic effects found in $H_2O$ on the conformation. For all values of *f*, conformation and structure are close to the one in $H_2O$ for *f* = 1. Minute differences are found at low q (less aggregation in DMSO than in $H_2O$ as revealed by the scattered intensity at zero angles) and at large q (local structure more helical in the rod-like length range). But the main behaviour, in the semi-dilute regime, for PSS at all *f* in DMSO is the same as for a classical hydrophilic polyelectrolytes in water. This is observed for the total scattering $S_T(q)$, both for the scattering profiles and for the abscissa of the maximum q*, which scales as $c_p^{0.44}$, not far from the $c_p^{0.5}$ dependency usually found for polyelectrolytes. Last but not least, this is **also observed** for the **single chain scattering, $S_1(q)$,** which matches precisely the form factor of a wormlike chain instead of a pearl necklace. Moreover, both for $S_T(q)$ and $S_1(q)$, there is in most of the cases no effect of charge rates *f*. Therefore we find complete agreement with the Manning-Oosawa condensation law (above the condensation threshold).

Moreover, the two structural studies support each other. From a global point of view, the **respective influences of THF addition and use of DMSO** follow the same trend, i.e. replacing water, which is a selective solvent for the sulfonated segments, by a non selective solvent. For DMSO, $S_1(q)$ shows definitely that we erase completely the pearl necklace structure and return towards the string-like one. Comparison of THF addition with use of DMSO shows that $S_T(q)$ behaves very similarly in the two cases. For the intermediate value of $f \sim 0.62$, 5% THF and DMSO are equivalent: $S_T(q)$ evolves towards the scattering measured for f = 1 in water. This suggests that in both cases the chain "feels" an **average effect** of the solvent. However this is not a universal proof: many complex behaviours may occur in other solvents. For the lower $f$, the change is still complete for DMSO, while the effect is only partial for THF, at least with 25% only. This again suggests that there is no solvent adsorption, and that the average quality of the solvent is the main parameter, as in the theoretical models. Adding more THF than done here could be successful. This also opens the way to the tuning of the pearl size.

As an extension of this work, it would be interesting to confirm our assumption concerning the increase of this effective charge of PSS with THF addition by measurements with a convenient technique. Also, kinetic structure investigations by SAXS of the PSS solutions at different scales immediately after dissolution in the mixture of $H_2O$ and THF, could enlightened the question of whether pearl necklace of PSS in $H_2O$ are at equilibrium or in a metastable state.

**Acknowledgments**. FB thanks J.P. Cotton and A. Brûlet for their help and support, and Ministère de la Recherche et de la Technologie for the 2.5 years funding of M.N. Spiteri Ph.D (full text available at www-llb.cea.fr).

**V-REFERENCES :**

(1) De Gennes, P.G.; Pincus, P.; Velasco, R. M.; Brochard, F. *J. Phys.(Paris)* **1976**, 37**,**1461-73.

(2) Dobrynin, A. V.; Colby, R. H. and Rubinstein, M. *Macromolecules* **1995**, 28,1859-71.

(3) Barrat, J. L.; Joanny J. F. *in Advances in Chemical Physics*; Prigogine, I., Rice, S.A., Eds.; Wiley:New York, **1996**; vol.X C IV, and references therein.

(4) Dobrynin, A. V.; Rubinstein, M.; Obukhov, S. P. *Macromolecules* **1996**, 29, 2974-9.

(5) Kantor, Y.; Kardar, M. *Europhys.Lett.* **1994**, 27, 643-8.

(6) a Dobrynin, A.V.; Rubinstein, M. *Macromolecules* **1999**, 32, 915-22. b Dobrynin, A.V.; Rubinstein, M. *Macromolecules* **2001**, 34,1964-1972.

(7) Micka, U.; Holm, C.; Kremer, K. *Langmuir* **1999**, 15, 4033-44.

(8) Limbach H.J.; Holm, C.; Kremer, K. *Europhys. Lett.* **2002**, 60, 566-72.

(9) Limbach H.J.; Holm, C. *J. Phys. Chem. B.* **2003**, 107, 8041-55.

(10) Schweins, R; Huber, K. *Macromol Symp.* **2004**, 211, 25.

(11) Liao, Q.; Dobrynin, A.V.; Rubinstein, M. *Macromolecules* **2006**, 39, 1920-1938.

(12) Williams, C.E. *Electrostatic Effects in Soft Matter and Biophysics*, edited by Holm C., Kekicheff P. and Podgornik R., NATO Science Series., vol. 46 (Kluwer Academic Publishers, Dordrecht) **2001**, 487-586.

(13) Baigl, D.; Sferrazza, M.; Williams, C.E. *Europhys. Lett.* **2003**, 62, 110-6.

(14) Essafi, W.; Lafuma F.; Williams, C. E. *in Macro-ion Characterization. From Dilute solutions to complex fluids*, K.S. Schmitz, ed., ACS Symposium Series 548, **1994**; 278-86.

(15) Essafi, W.; Lafuma, F.; Williams, C.E. *Eur. Phys. Journal B*, **1999**, 9, 261-6.

(16) Spiteri, M.N; Ph.. D, Orsay, **1997** and Spiteri, M.N; Williams, C.E; Boué, F. *Macromolecules* **2007**, 40, 6679.

(17) Essafi, W.; Lafuma, F.; Williams, C.E. *J. Phys. II* **1995,** 5, 1269-75.

(18) Essafi, W.; Lafuma, F.; Baigl, D.; Williams, C.E. Europhys. Lett. **2005**, 71, 938-44.

(19) Aseyev, V.O.; Tenhu, H.; Klenin, S.I. *Macromolecules* **1998**, 31, 7717-22.

(20) Aseyev, V.O.; Klenin, S.I.; Tenhu, H.; Grillo, I.; Geissler, E. *Macromolecules* **2001**,34, 3706-9.

(21) Schweins, R.; Lindner, P.; Huber, K. *Macromolecules* **2003**, *36*, 9564

(22) Goerigk, G.; Schweins, R; Huber, K.; Ballauf, M. *Europhys. Lett.* **2004** *66*, 331-337

(23) Kiryi, A.; Gorodyska, G.; Minko, S.;Jaeger, W.; Stepanek, P.; Stamm M. *JACS* **2002**, *124*, 13454-13462.

(24) Kirwan, L. J.; Papastravou, G.; Borkovec M. *Nanoletters* **2004** *4* 149-152.

(25) Gromer, A.; Rawiso, M.; Maaloum, M. *Langmuir*, **2008**, *24*, pp 8950–8953

(26) Nierlich, M. ; Williams, C.E.; Boué, F; Cotton, J.P.; Daoud, M.; Farnoux, B.; Jannink, G.; Picot, C.; Moan, M.; Wolff, C.; Rinaudo, M.; de Gennes P.G. *J. Phys. France* **1979**, 40, 701-4.

(27) Nierlich, M.; Boué, F.; Lapp, A.; Oberthür, R. *Coll.Pol.Sci.* **1985**, 263, 955.

(28) Jannink, G. *Makromol. Chem., Macromol. Symp.* **1986**, 1, 67-80.

(29) Heitz, C.; Rawiso, M.; Francois, J. *Polymer* **1999**, 40,1637.

(30) Oosawa F., *Polyelectrolytes*; M.Dekker, New York: New York **1971**.

(31) Manning G. S. *J. Chem. Phys.* **1969**; 51, 924-933 and 934-8.

(32) Combet, J.; Isel, F.; Rawiso, M.; Boué, F. *Macromolecules* **2005**, 38, 7456-7469.

(33) Baigl, D.; Ober R.; Dan Qu, D.; Fery, A.; Williams, C. E. *Europhys. Lett*. **2003**, 62, 588 - 594.

(34) Dan Qu, D.; Baigl, D.; Williams, C.E.; Möhwald, H.; Fery, A. *Macromolecules* **2003**,36, 6878-83.

(35) Alexei Chepelianskii, Farshid Mohammad-Rafiee, Elie Raphael On the effective charge of hydrophobic polyelectrolytes arxiv.org/abs/0710.2471.

(36) Boué, F.; Cotton, J.P.; Lapp, A; Jannink, G. *J. Chem. Phys* **1994,** 101, 2562-8.

(37) Jousset. S, Bellissent H, Galin J.C, *Macromolecules*, **1998**, 31, 4520-30.

(38) Waigh, T.A.; Ober, R.; Williams, C.E.; Galin, J.C. *Macromolecules*, **2001**, 34, 1973-80.

(39) Uyaver, S.;Seidel, C.J. *J.Phys.Chem.B* **2004**, 108,18804.

(40) Vallat P.; Schossler F.; Cathala J.M.; Rawiso M. *Euro. Phys. Lett.*, **2008**, 82 (2)

(41) Spiteri, M.N.; Boué, F.; Lapp, A.; Cotton, J.P. *Physical Review Letters* **1996**, 77, 26, 5218-5220.

(42) Makowski, H. S; Lundberg, R. D; Singhal, G. S. *U.S Patent* 3 870 841, **1975** to Exxon Research and Engineering Company.

(43) Baigl, D.; Seery, T. A. P.; Williams, C. E. *Macromolecules* **2002**, 35, 2318-26.

(44) Vink, H. *Makromol. Chem*. **1980**, *182*, 279.

(45) Mc Cormick C. L.; Chen G. S., *J. Polym. Sci. Polym. Chem. Ed.* **1982**, 20, 817-38.

(46) Heinrich, M; PhD, Louis Pasteur University, Strasbourg, **1998**.

(47) Dubois, E.; Boué, F. *Macromolecules* **2001**, *34,* 3684-3697.

(48) Heinrich, M.; Rawiso M.; Zilliox, J.G.; Lesieur P., Simon, J.P. *Eur. Phys. J. E.* **2001**, *4*,131.

(49) Zhang, Y. ; Douglas, J.F.; Ermi, B. D. ; Amis, E.J. *J. Chem. Phys.* **2001**, 114, 3299-3313.

(50) Cotton J.P. Comment faire une calibration absolue des mesures de DNPA; LLB Web site, www-llb.cea.fr

(51) Herzfeld, J.; Olbris, D. *ENCYCLOPEDIA OF LIFE SCIENCES* **2002** Macmillan Publishers Ltd.

(52) Sharp, P; Bloomfield, V.A. *Biopolymers* **1968,** 6, 1201.

(53) Des Cloizeaux, J. *Macromolecules* **1973,** 6, 403.

(54) Van der Maarel, J. R.C.; Groot, L. C. A.; Hollander, J. G.; Jesse, W.; Kuil, M. E.; Leyte, J. C.; Leyte-Zuiderweg, L. H.; Mandel, M.; Cotton J. P.; Jannink, G.; Lapp, A.; Farago, B. *Macromolecules* **1993,** 26, 7295 – 7299.

**Figure Captions :**

Figure 1: The chemical structure of the polyelectrolytes used in this work. f is the rate of sulfonation, or "chemical charge".

Figure 2: Evolution of the SAXS profiles with the percentage of added THF to aqueous solutions of AMAMPS. (a) $f = 0.40$. (b) $f = 0.60$. The polyelectrolyte concentration $c_p = 0.32$ M/L.

Figure 3: Evolution of the SAXS profiles with the percentage of added THF to aqueous solutions of PSSNa. (a) PSSNa $f = 1$. (b) PSSNa $f = 0.58$. (c) PSSNa $f = 0.38$. The polyelectrolyte concentration $c_p = 0.32$ M/L.

Figure 4: Evolution of q*, I(q*), and I(q $\rightarrow$ 0) as a function of THF addition, for PSSNa $f = 0.38$ and $c_p = 0.32$ M/L.

Figure 5: Evolution of the total scattering function per monomer $S_T(q)/c_p$ for h-PSSNa at $f=0.64$ in d-DMSO as a function of polymer concentration.

Figure 6 : Log-log representations of the intrascattering function $S_1(q)/c$ (PSSNa $f=0.64$) in DMSO as a function of polymer concentration.

Figure 7: Fit (solid line) of the SANS data (symbols) according to the Sharp and Bloomfield model in Kratky representation for $c_p = 0.085$ M/L and 0.85 M/L.

Figure 8: Universal des Cloiseaux representation of $S_1(q)$ for the concentrations $c_p$ as in Figure 6.

Figure 9: Log log representation of the persistence length $l_p$ and the radius of gyration $R_g$ as a function of the ionic strength of the solution $I_S = fc_p+2c_s$; $c_s$ is the concentration of the added salt concentration, for PSSNa $f =0.64$ in DMSO.

Figure 10: Comparison of the intrascattering function $S_1(q)$ of the fully charged PSSNa $f=1$ in aqueous solvent and that of PSSNa $f=0.64$ in organic solvent : DMSO, $c_p = 0.34$ M/L (Log-log representation).

Figure 11: Evolution of the apparent structure factor $S_T/S_1(q)$ vs q as a function of polyelectrolyte concentration $c_p$. The insert is a log-log plot of the variation of $q_S$ (the peak abcissa of the apparent structure factor), q* (the peak abcissa of the total structure function) = f($c_p$).

Figure 12: Evolution of the total scattering function as a function of the charge rate of PSSNa in DMSO (a) deuterated PSSNa in hydrogenated DMSO at $c_p= 0.17$ M/L, (b) for $c_p= 0.34$ M/L and (c) hydrogenated PSSNa in deuterated DMSO at $c_p= 0.34$ M/L.

Figure 13: Log-log plots of the intrascattering function $S_1(q)/c_p$ for polyions of different charge rate $f$ at a polymer concentration $c_p$= 0.34 M/L, the insert is the Kratky plot $q^2S_1(q)/c_p$ of the intrachain scattering function $S_1(q)$ of polyions, measured at a polymer concentration $c_p$ = 0.34 M/L, for different degree of charge rates $f$.

Figure 14: Comparison of the persistence lengths as a function of the ionic strength $I_S$ = $fc_p+2c_s$ in water and in DMSO ( logarithmic representation).

Figure 15: The apparent structure factor = $S_T(q)/S_1(q)$ for the three sulfonation rates $f = 1$, 0.64, and 0.36 at a polymer concentration $c_p = 0.34$ M/L.

PSS

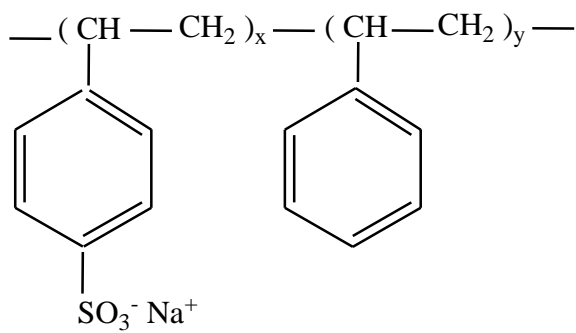

*f=x/x+y*

AMAMPS

*f=x/x+y*

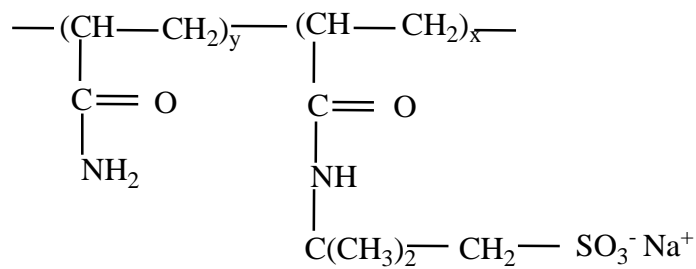

$(C_3ON H_5)_{(1-f)}$- $(C_3ON H_4 –C_4H_8SO_3Na)_f$ =71(1-f) + (70 + 159)f = 71 + 158. f

Figure 1

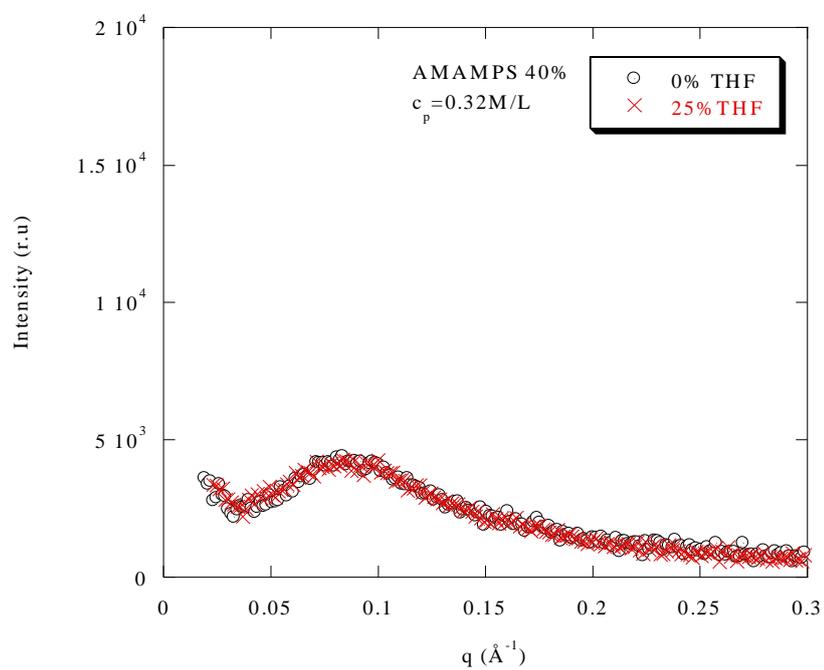

(a)

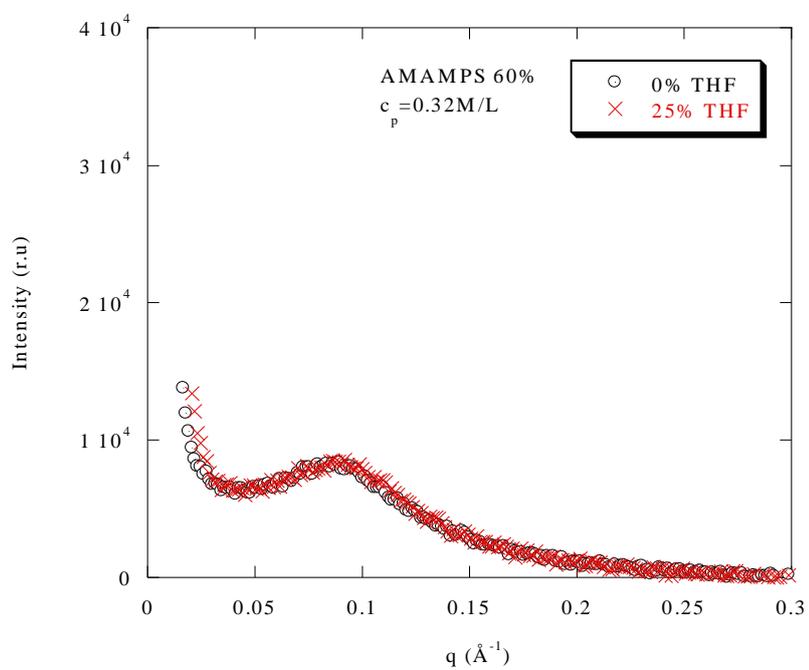

(b)

Figure 2

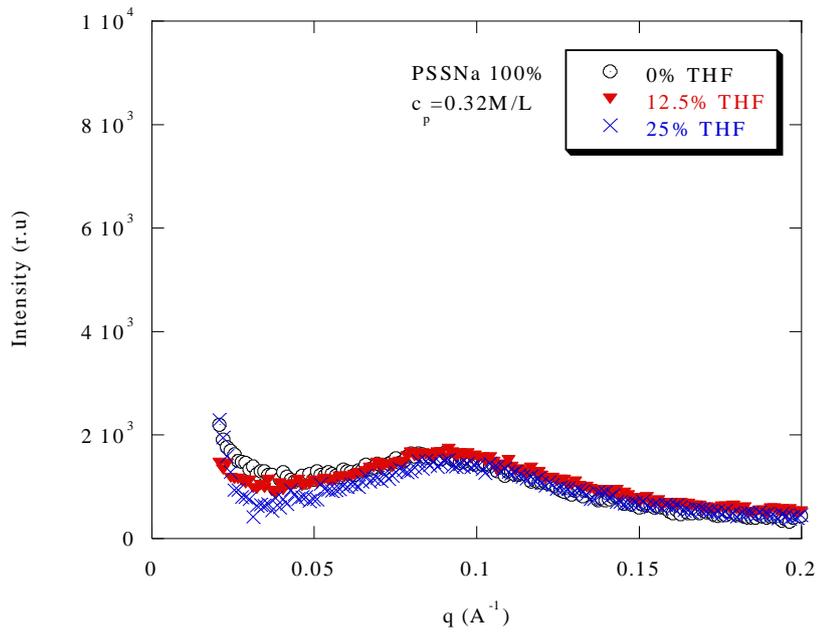

(a)

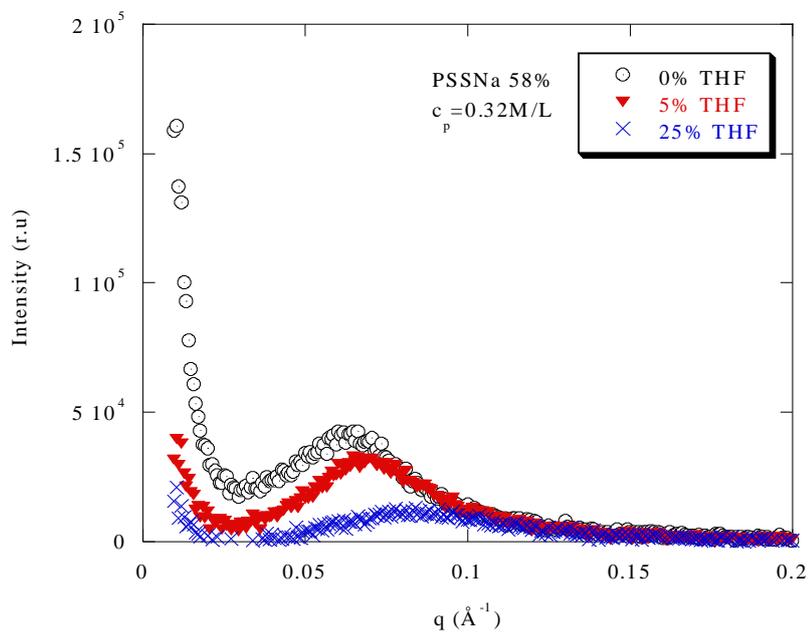

(b)

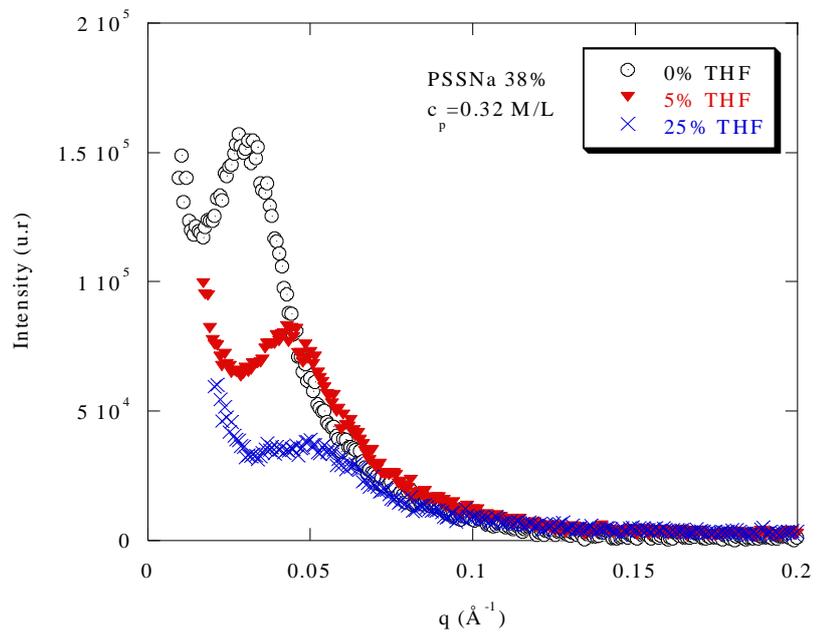

(c)

Figure 3

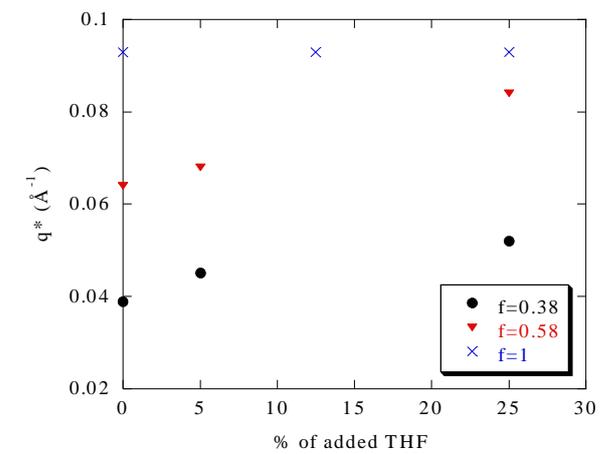

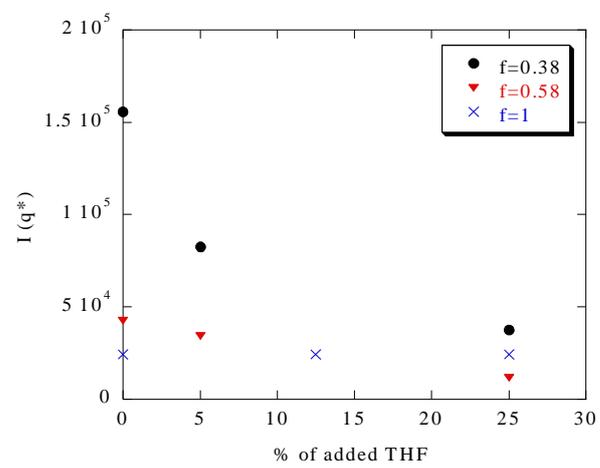

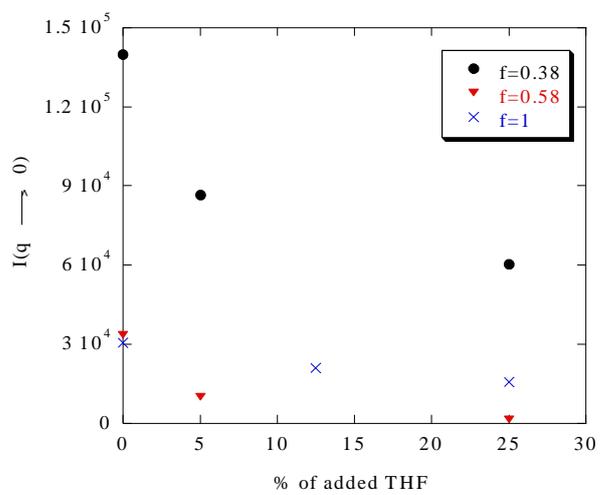

Figure 4

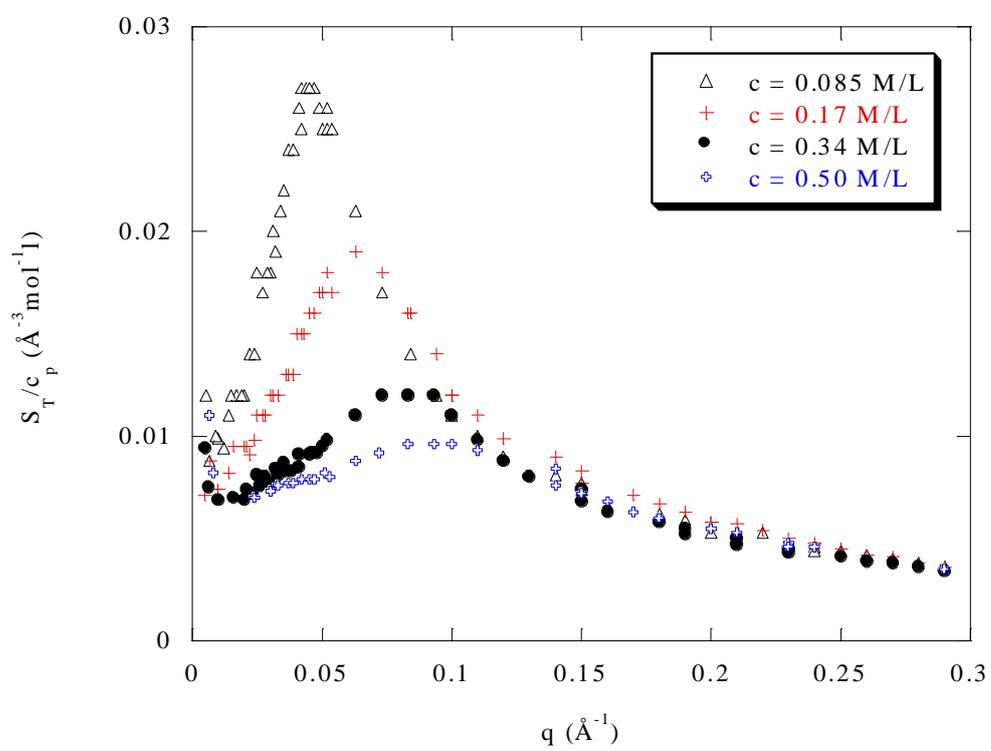

Figure 5

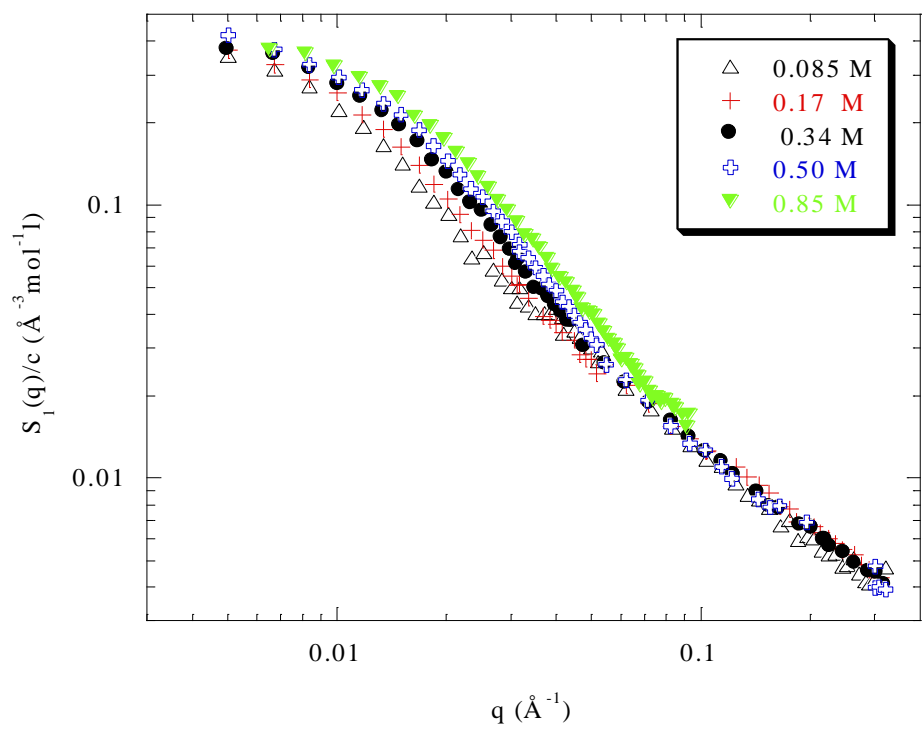

Figure 6

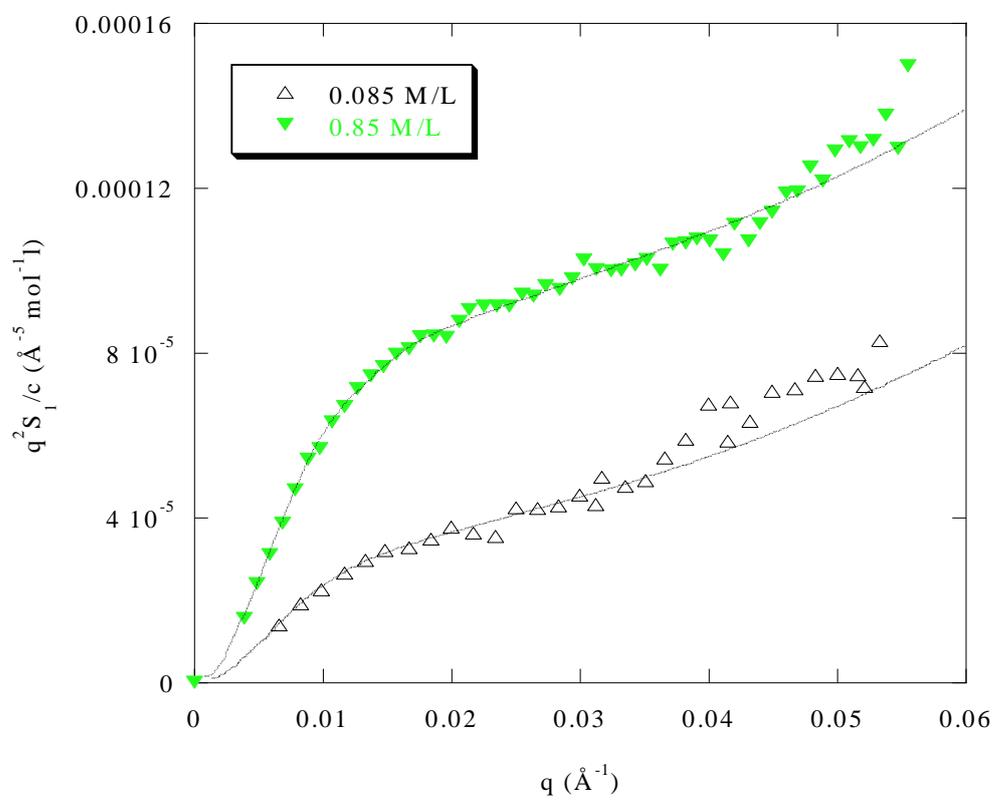

Figure 7

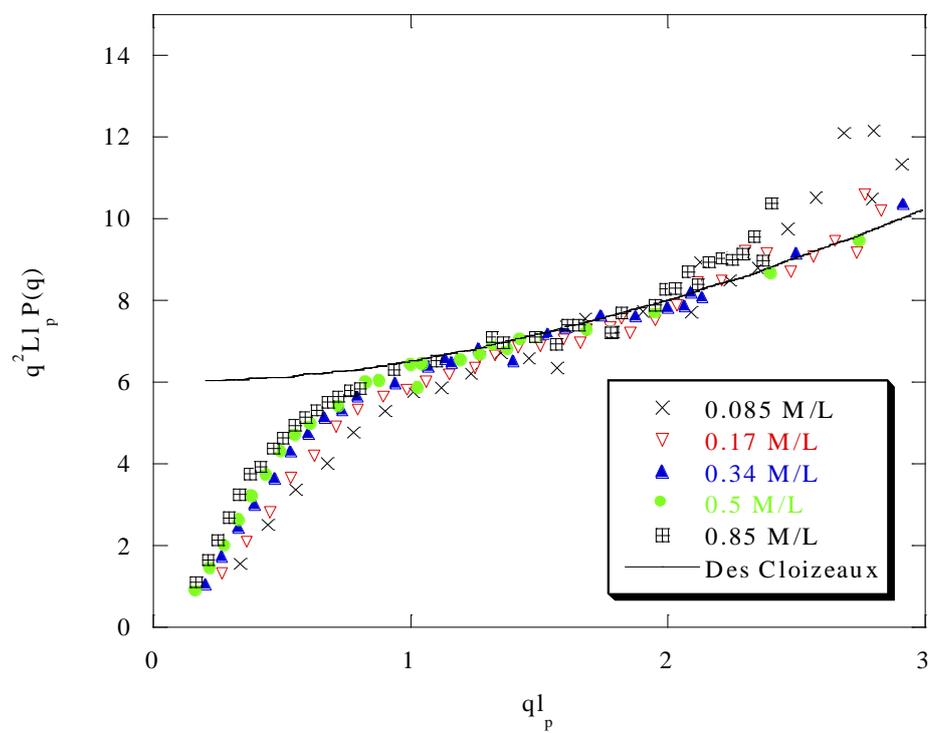

Figure 8

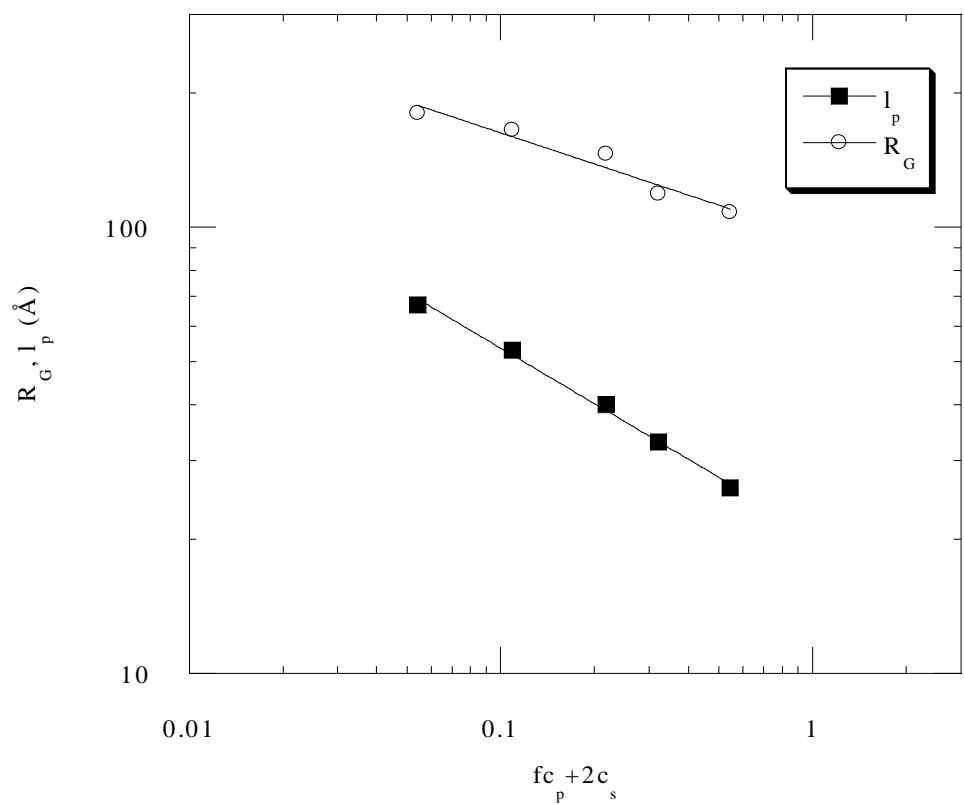

Figure 9

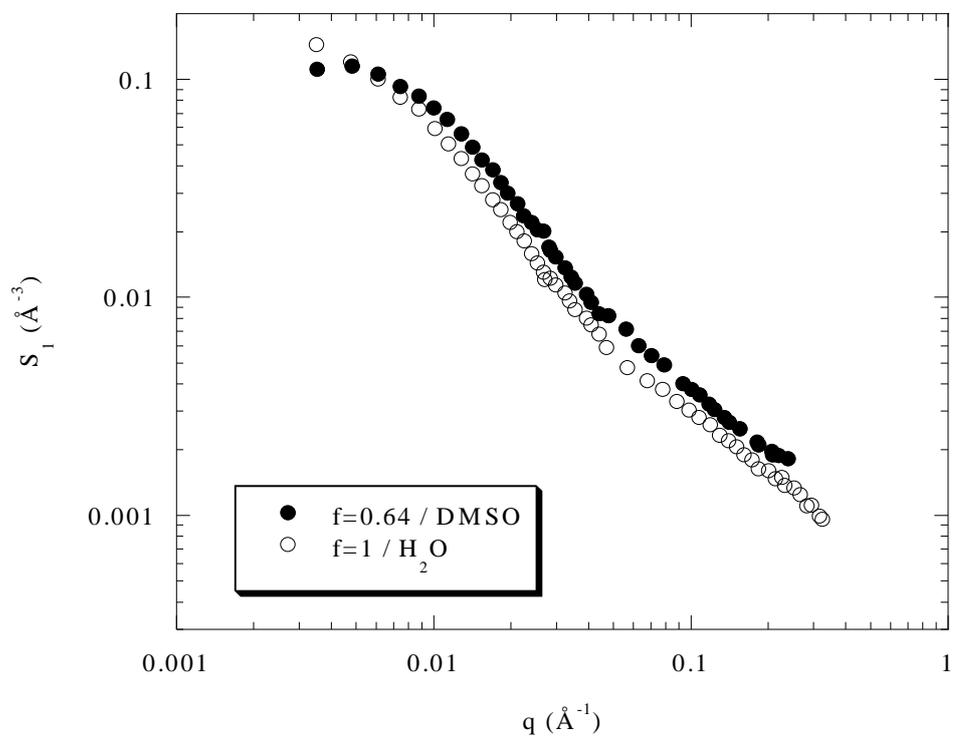

Figure 10

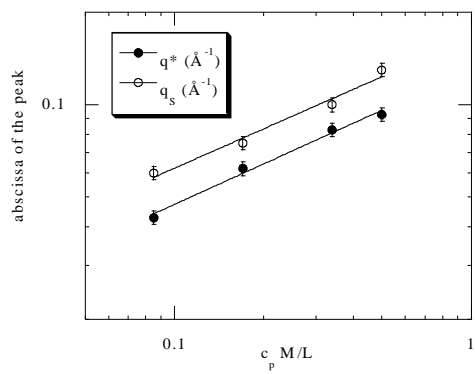

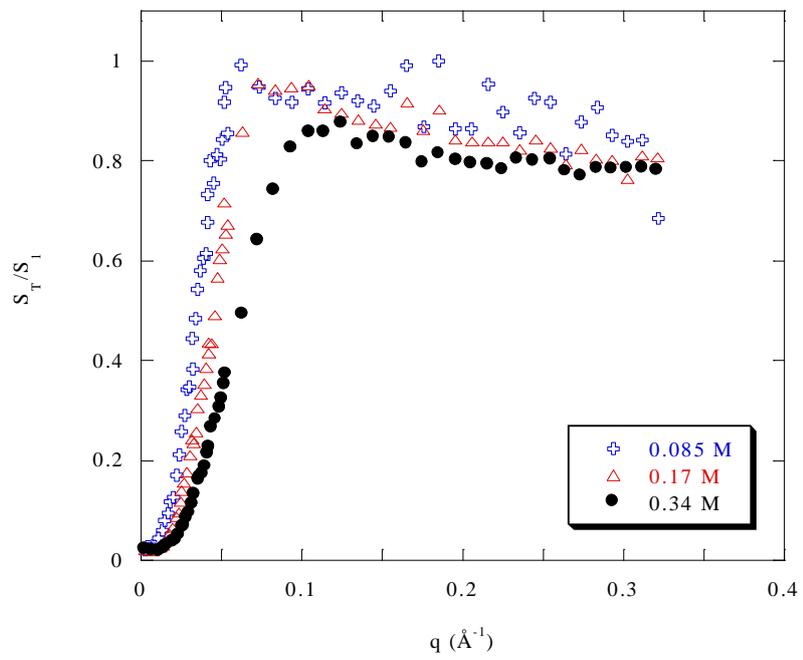

Figure 11

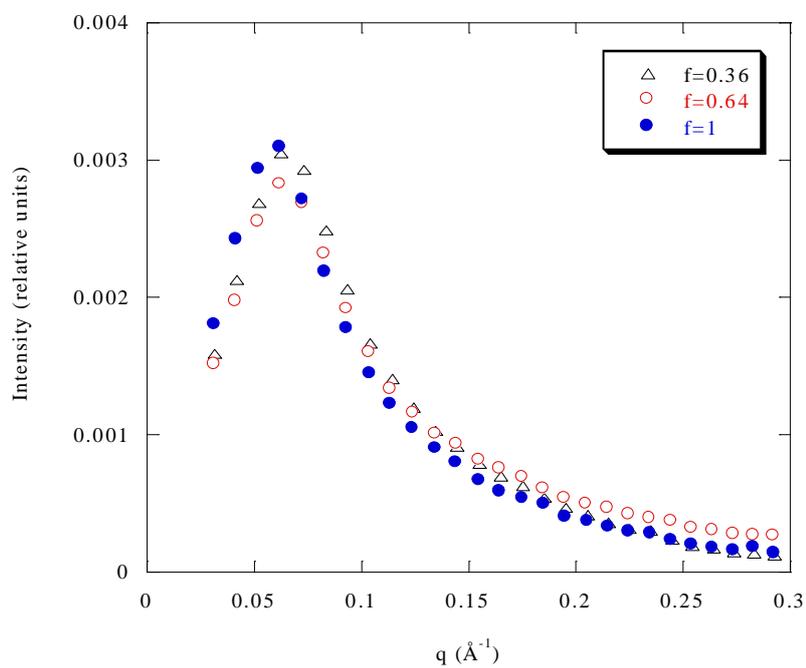

(a)

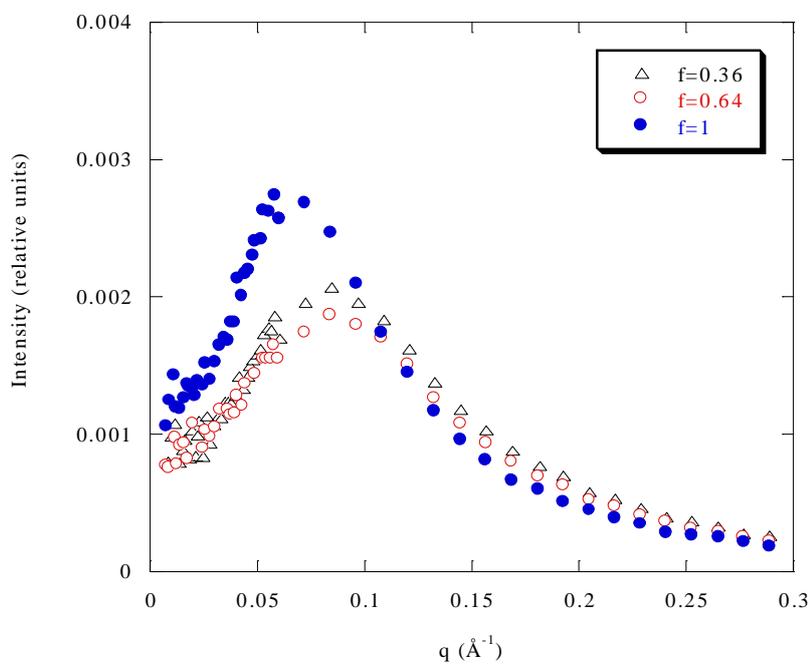

(b)

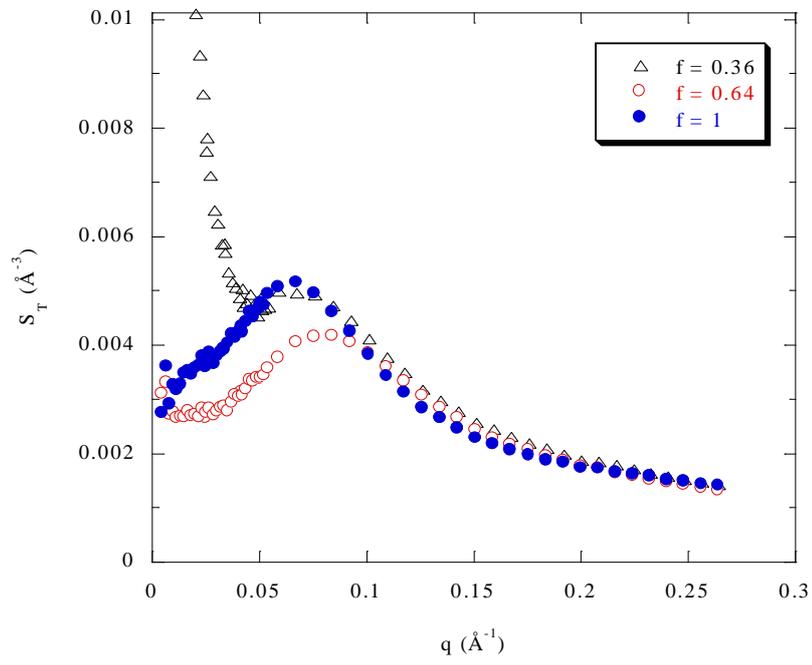

(c)

Figure 12

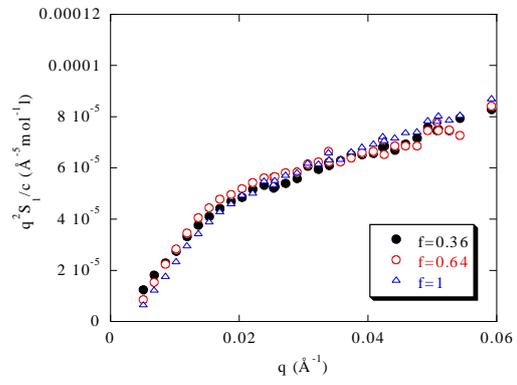

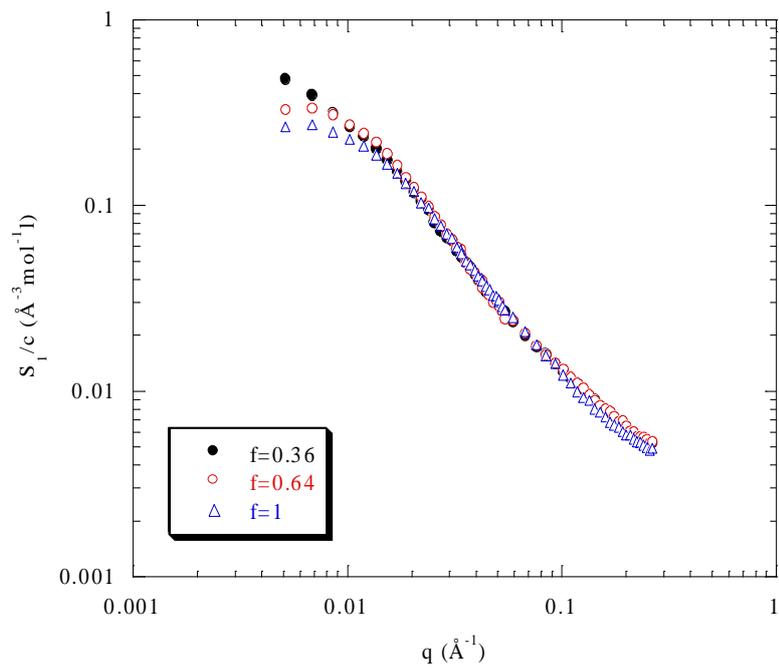

Figure 13

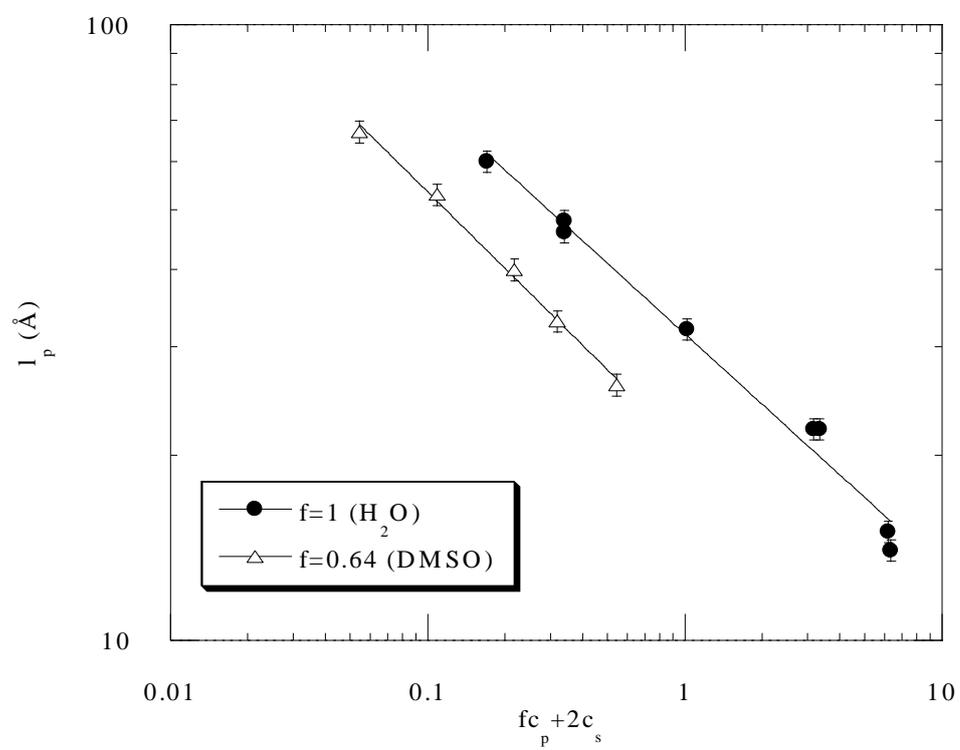

Figure 14

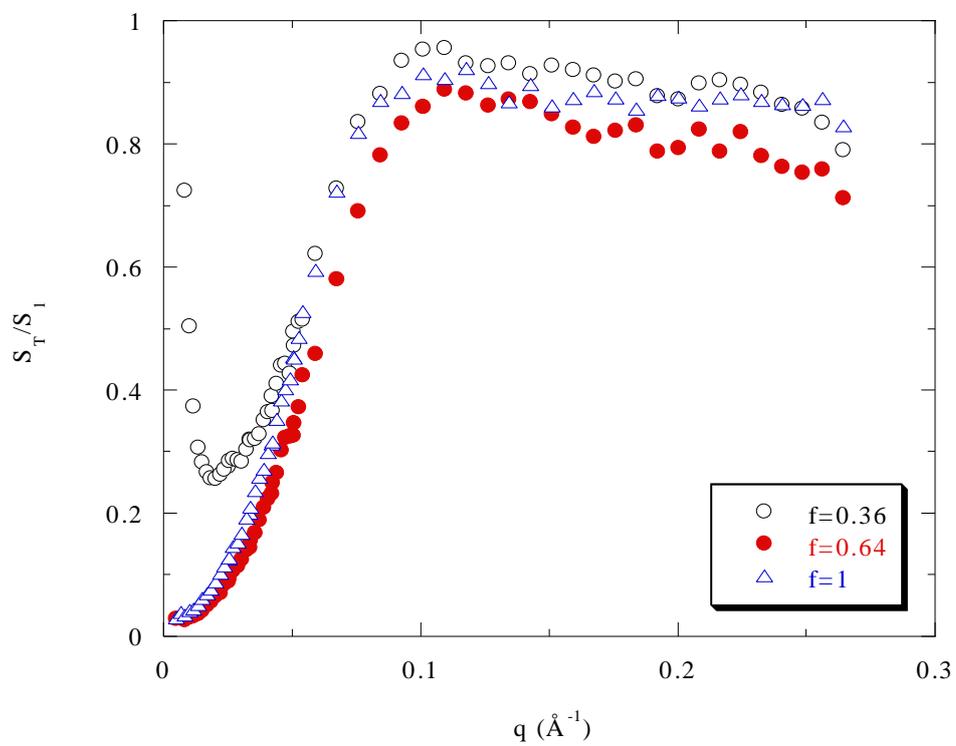

Figure 15

**Table 1 : Characteristics of the used Polymers**

| Technique/ Polyelectrolyte | $M_0$ g/mol | $V_{mol}$ (cm$^3$) | $M_w$ (g/mol) | $M_w/M_n$ | $N_w$ |
|---|---|---|---|---|---|
| SAXS/h-PS | 104 | 98 | 250000 | 2 | 2403 |
| SAXS/h AMAMPS  f=1 | 71+f*158 | | 650,000 | 2.6 | 2838 |
| f=0.6 | | | 471,000 | | |
| f=0.4 | | | 381,000 | | |
| SANS/h-PS | 104 | 98 | 67500 | 1.03 | 625 |
| SANS/d-PS | 112 | 98 | 73000 | 1.04 | 652 |
| SANS/h-PSSNa (f=1) | 206 | 108 | 150000 | 1.12 | 730 |
| SANS/d-PSSNa (f=1) | 213 | 108 | 170000 | 1.2 | 800 |
| SANS d f=0.36 ± 2%. | 148.4 | 101 | 107 000 | | 720 |
| h f=0.36 ± 2%. | 140.7 | 101 | 101 000 | | 720 |
| SANS d f=0.64± 1% | 176.6 | 105 | 127 000 | | 720 |
| h f=0.64± 1% | 169.3 | 105 | 122 000 | | 720 |

**Table 2 : Characteristics of the DMSO solvent**

| Solvent | formula | permittivity $\varepsilon$ | $V_{mol}$ (cm$^3$) | $M_0$ (g/mol) | Scattering length (x10$^{-12}$ cm) |
|---|---|---|---|---|---|
| H-DMSO | SO(CH$_3$)$_2$ | 46 | 71.0 | 78.1 | -0.054 |
| D-DMSO | SO(CD$_3$)$_2$ | 46 | 71.0 | 84.1 | 6.192 |

**Table 3 : Values of the different contrast lengths of polyions in DMSO**

| | PSSNa f=1 | PSSNa f=0.64 | PSSNa f=0.36 |
|---|---|---|---|
| $V_{mol}$ PSS (cm$^3$) | 115 | 109.5 | 103.5 |
| $x_{ZAC}$ | 0.84 | 0.81 | 0.80 |
| $|k|_{ZAC}$ (x 10$^{-12}$ cm) | 3.600 | 3.858 | 3.934 |
| $k_{ST}$ (x 10$^{-12}$ cm) | -5.30 | -5.68 | -5.83 |

**HYDROPHOBIC POLYELECTROLYTES IN BETTER POLAR SOLVENT.
STRUCTURE AND CHAIN CONFORMATION AS SEEN BY SAXS AND SANS**

Wafa ESSAFI, Marie-Noelle SPITERI, Claudine WILLIAMS and François BOUE